\documentclass[onecolumn]{aastex631}

\usepackage{txfonts}
\usepackage{color}
\usepackage{hyperref}
\usepackage{graphicx}
\usepackage{subfigure}
\usepackage{float}

\usepackage{threeparttable}

\hypersetup{hypertex=true,
	colorlinks=true,
	linkcolor=blue,
	anchorcolor=blue,
	citecolor=blue}
\hypersetup{colorlinks=true,
	linkcolor=blue,
	filecolor=blue,      
	urlcolor=blue,
	citecolor=blue}

\submitjournal{ApJ}
\shorttitle{Search for and analysis of eclipsing binaries}
\shortauthors{Wang et al.}

\begin{document}
	
	\title{Search for and analysis of eclipsing binaries in the LAMOST Medium-Resolution Survey field. I. RA: $\textbf{23}^h$$\textbf{01}^m$$\textbf{51}^s$, Dec: +34$^\circ$36$^\prime$45$^{\prime \prime}$}
	
	\correspondingauthor{Kai Li}
	\email{kaili@sdu.edu.cn}
	
	\author{Jing-Yi Wang}
	\affil{Shandong Key Laboratory of Optical Astronomy and Solar-Terrestrial Environment, School of Space Science and Technology, Institute of Space Sciences, Shandong University, Weihai, Shandong, 264209, China}
	
	\author{Kai Li}
	\affil{Shandong Key Laboratory of Optical Astronomy and Solar-Terrestrial Environment, School of Space Science and Technology, Institute of Space Sciences, Shandong University, Weihai, Shandong, 264209, China}
	
	\author{Xiang Gao}
	\affil{Shandong Key Laboratory of Optical Astronomy and Solar-Terrestrial Environment, School of Space Science and Technology, Institute of Space Sciences, Shandong University, Weihai, Shandong, 264209, China}
	
	\author{Di-Fu Guo}
	\affil{Shandong Key Laboratory of Optical Astronomy and Solar-Terrestrial Environment, School of Space Science and Technology, Institute of Space Sciences, Shandong University, Weihai, Shandong, 264209, China}
	
	\author{Li-Heng Wang}
	\affil{Shandong Key Laboratory of Optical Astronomy and Solar-Terrestrial Environment, School of Space Science and Technology, Institute of Space Sciences, Shandong University, Weihai, Shandong, 264209, China}
	
	\author{Dong-Yang Gao}
	\affil{Shandong Key Laboratory of Optical Astronomy and Solar-Terrestrial Environment, School of Space Science and Technology, Institute of Space Sciences, Shandong University, Weihai, Shandong, 264209, China}
	
	\author{Ling-Zhi Li}
	\affil{Shandong Key Laboratory of Optical Astronomy and Solar-Terrestrial Environment, School of Space Science and Technology, Institute of Space Sciences, Shandong University, Weihai, Shandong, 264209, China}
	
	\author{Ya-Ni Guo}
	\affil{Shandong Key Laboratory of Optical Astronomy and Solar-Terrestrial Environment, School of Space Science and Technology, Institute of Space Sciences, Shandong University, Weihai, Shandong, 264209, China}
	
	\author{Xing Gao}
	\affil{Xinjiang Astronomical Observatory, 150 Science 1-Street, Urumqi 830011, China}
	
	\author{Guo-You Sun}
	\affil{Xingming Observatory, Urumqi, Xinjiang, China}
	
	\begin{abstract}
		
		Eclipsing binaries (EBs) play an important astrophysical role in studying stellar properties and evolution. By analyzing photometric data in the LAMOST Medium-Resolution Survey field, RA: $23^h$$01^m$$51.00^s$, Dec: +34$^\circ$36$^\prime$45$^{\prime \prime}$, 48 EBs are detected and 2 are newly discovered. This specific field has been observed 52 times by the LAMOST Medium-Resolution Survey DR 9, which facilitates a comprehensive analysis of the EBs. For EBs with LAMOST medium-resolution spectra, radial velocity curves were obtained, and their precise orbital parameters were determined by simultaneously analyzing photometric light curves and radial velocity curves. For the other EBs with only photometric light curves, we used the q-search or the temperature ratio method to determine their initial mass ratios and then determined the orbital parameters. It is found that 15 EBs belong to detached systems, 1 to semi-detached systems, and 32 to contact systems. Based on the O-C analysis for 26 EBs with sufficient eclipsing times, we found a long-term decrease in the orbital period of 11 EBs and a continuous increase of 5 EBs, which are due to the material transfer between the two components. The O-C curve of 1 EB shows a distinct periodic variation, which is caused by the light travel time effect, and the third body is likely to be a black hole. By applying the spectral subtraction method to 13 EBs with LAMOST medium-resolution spectra, 10 systems exhibit distinct H$\alpha$ emission lines, in which 1 system exhibits double-peaked lines near phases 0.25 and 0.75, implying strong chromospheric activity. In the mass-luminosities and mass-radius distributions, most of the more massive components are less evolved than the less massive ones.
		
	\end{abstract}
	
	
	\section{Introduction} \label{sec1}
	
	Variable stars are stars whose brightness varies over time, either due to intrinsic physical factors, such as pulsations and eruptions, or due to extrinsic factors, such as eclipses \citep{a1}. Among variable stars, EBs play an important astrophysical role. 
	The study of EBs is crucial for the precise determination of the fundamental physical parameters of stars, which are foundational for understanding the structure of stars and can verify stellar evolution theories \citep{a2}. EBs are used as a standard candle to determine the distances of various celestial bodies, including the Magellanic Clouds, M31, and M33 \citep{a130,a131,a30,a133}. Studying EBs offers opportunities for searching for black holes and exoplanet candidates \citep{a115}. In binary systems, when two components are close enough, material and energy transfer between the two components affects evolution, and these systems are called close binaries. Based on the geometrical structure of close binaries, \cite{a86} classified them into three categories: detached binaries where neither component fills the Roche Lobe, semi-detached binaries where one component fills the Roche Lobe and the other does not, and contact binaries where both components fill the Roche Lobe. Contact binaries are usually believed to form from detached binaries through angular momentum loss due to magnetic braking and gravitational radiation, and they may merge to form blue stragglers and FK Comae stars \citep{a88,a89,a90,a141}.

	Photometric and spectroscopic observations are important tools for determining physical parameters and analyzing the structure and evolution of EBs. Photometric light curves can be used to obtain physical parameters, such as photometric mass ratio, orbital inclination, relative radii, and luminosity ratio. Reliable photometric mass ratios can be obtained for semi-detached and contact binaries with the total eclipse minimum \citep{a31,a70}. Additionally, the eclipsing times can be used to analyze orbital period variations, contributing to the study of dynamical evolution and searching for additional companions. Spectroscopic observations can be used to obtain the radial velocity curves, thereby precisely determining the mass ratio and mass function of EBs. Stellar spectra can also be used to derive atmospheric parameters, as well as to analyze chromospheric activity \citep{a125,a138,a139}. The integrated analysis of photometric light curves and radial velocity curves enables the determination of accurate physical parameters of EBs, especially their absolute physical parameters \citep{a124}.

	The phenomenon of magnetic activity in binary systems has been frequently observed and studied (e.g., \citeauthor {a79} \citeyear{a79}; \citeauthor {a80} \citeyear{a80}; \citeauthor {a81} \citeyear{a81}; \citeauthor {a82} \citeyear{a82}; \citeauthor {a143} \citeyear{a143}). This activity is triggered by the heating of the magnetic field in the outer atmosphere, which in turn leads to a variety of observable phenomena such as chromospheric activity, starspots, and activity cycles \citep{a118}. There are many spectral lines that can serve as indicators of chromospheric activity, including the Balmer series (H$\alpha$: 6562.8 Å, H$\beta$: 4861 Å, H$\gamma$: 4340.5 Å, H$\delta$: 4102 Å), the Ca II H \& K lines (3968, 3933 Å), and the Ca II infrared triplet (IRT; 8498, 8542, 8662 Å). Among these, the H$\alpha$ spectral line and the Ca II H \& K lines, which are easily observable, are the most commonly utilized for studying chromospheric activity \citep{a59,a60}.
	Another phenomenon of magnetic activity that has been extensively studied is the O’Connell effect \citep{a37},  which has been observed on binaries (e.g., \citeauthor {a85} \citeyear{a85}; \citeauthor {a93} \citeyear{a93}; \citeauthor {a92} \citeyear{a92}). O'Connell effect is manifested by that the two out-of-eclipse maxima of the light curves are not equal. In general, there are several explanations for the O'Connell effect: starspots due to magnetic activity \citep{a94}, accretion of material between two components \citep{a95}, circumstellar material around the binary \citep{a84}, and asymmetry of material due to the coriolis force \citep{a97}, etc. Of these, starspots are considered as the most likely cause of the O'Connell effect \citep{a98,a110,a99}.
	
	We implemented a project called the “$\emph{L}$AMOST $\emph{E}$clipsing $\emph{B}$inaries $\emph{S}$earch (LEBS) Project", which aims to search for and analyze EBs in the LAMOST Medium-Resolution Sky Survey field. This paper is the first work of our project, searching for and analyzing EBs around RA: $23^h$$01^m$$51^s$, Dec: +34$^\circ$36$^\prime$45$^{\prime \prime}$. We selected this field as the target field because it has been observed 52 times by the LAMOST Medium-Resolution Survey DR9\footnote{\url{https://www.lamost.org/dr9/v1.0/}}. This provides an opportunity for obtaining a complete phase coverage of the radial velocity curve, and the spectra facilitate a comprehensive analysis of EBs. Section 2 describes observations and data reduction. Orbital parameters were determined in Section 3. Orbital period variations were analyzed in Section 4. In Section 5, the chromospheric activities of 13 binaries were analyzed using LAMOST medium-resolution spectra. Section 6 describes the discussions and conclusions.

	\section{Observations and Data Reduction}
	\subsection{Photometry}
	\subsubsection{Ground-based photometric data}
	
	A 10 cm telescope at Xingming Observatory\footnote{\url{http://xjltp.china-vo.org/}} was used to obtain photometric data. The telescope is equipped with an APOGEE U16M CCD, possessing 4096 × 4096 square pixels and offering a field of view of 239$^\prime$× 239$^\prime$. During the observations, the filters of the Sloan r$^\prime$ and i$^\prime$ were used. The center of the telescope observed towards RA: $23^h$$01^m$$51^s$, Dec: +34$^\circ$36$^\prime$45$^{\prime \prime}$. The exposure time for both filters was 120 seconds, the readout time was 40 seconds for each image, and the final cadence was 320 seconds because the two filters were observed alternately. The observations were conducted for 13 days, from September 2 to September 22, 2021. More than 900 valid images were collected for each filter, which were the ones after visual exclusion of those impacted by obvious poor weather conditions.
	
	First, all observed CCD images were preprocessed, including bias subtraction, dark subtraction, and flat correction. The World Coordinate System information was acquired by matching the CCD frames with the Fourth U.S. Naval Observatory CCD Astrograph Catalog (UCAC4) using the \emph{Xparallax viu}\footnote{\url{https://www.xparallax.com/}} software. We used the \emph{Sep}\footnote{\url{https://sep.readthedocs.io/en/v1.1.x/}} package to detect sources. \emph{Photutils}\footnote{\url{https://photutils.readthedocs.io/en/stable/index.html}} package was used for aperture photometry to obtain light curves for all sources in the field of view. 10 sources with strong and constant brightness were selected as reference stars. After determining the reference stars, we obtained differential light curves for all sources. Since there are no confirmed standard stars in the field of view, using the same method as \cite{a87}, we carried out the visual magnitude calibration of sources by subtracting the average of the observed magnitudes from the visual magnitudes of the 10 reference stars. We first chose sources with large standard deviations in the light curves as variable star candidates. Next, we manually excluded non-variable stars with spuriously large standard deviations, which may have been caused by the images that were not completely eliminated due to poor weather conditions, or by the stars located at the edge of the field and those contaminated by satellite crossings. Finally, we identified 53 variable stars, 48 are EBs and 5 are pulsating variables. The EBs we detected were renamed as LEBS-01 $\sim$ LEBS-48. Among them, EBs with photometric data and LAMOST medium-resolution spectra were LEBS-01 $\sim$ LEBS-13, and EBs with only photometric data were LEBS-14 $\sim$ LEBS-48. By cross-matching with previously discovered EBs, we found that two EBs are newly discovered. The five pulsating variable stars have all been discovered previously, and the specific details of these stars are not discussed since it is beyond the scope of this paper. We used the period which provided by the The International Variable Star Index (VSX) website\footnote{\url{https://www.aavso.org/vsx/index.php}} for the 46 confirmed EBs. For 2 newly discovered targets, we used the Lomb-Scargle (LS) Periodograms \citep{a102} to obtain their periods. The light curves were folded according to the corresponding period. The Gaussian Process (GP) method is widely used for time series analysis and is often used to model light curves and identify outliers in EBs \citep{a126}. We employed the GP method to fit the light curves and calculated the standard deviations ($\sigma$) between the observed data and the GP model predictions. Data points with residuals exceeding 3$\sigma$ were considered as outliers, thereby yielding more accurate and reliable light curve models.
	
\begin{figure*}[ht!]
		\centering
		\includegraphics[width=5.5cm]{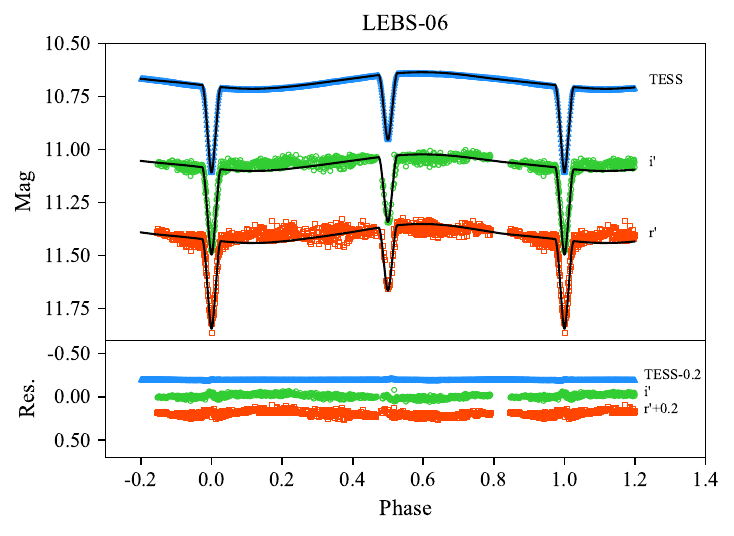}
		\includegraphics[width=5.5cm]{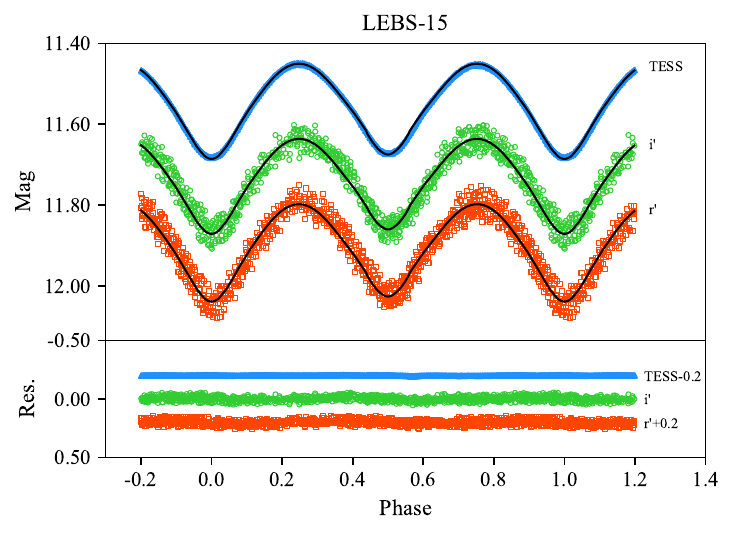}
		\includegraphics[width=5.5cm]{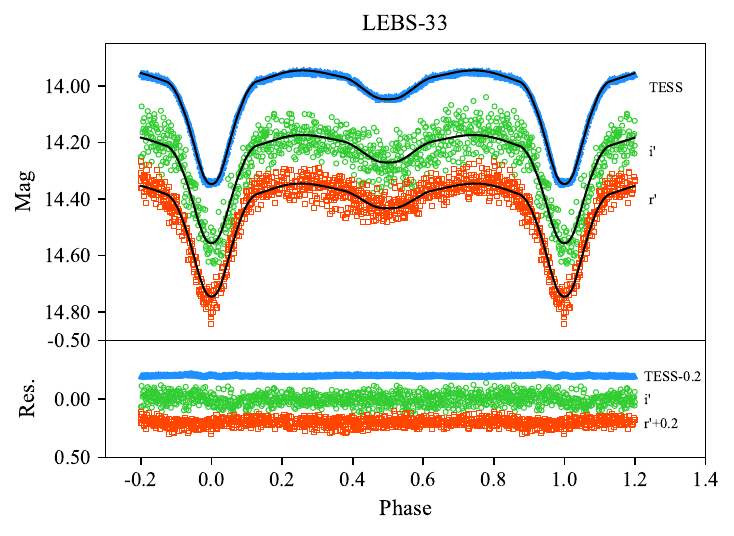}   
		\caption{Light curves for the 53 variable stars. The colored symbols represent observational light curves, and the black lines represent theoretical light curves. The light curves of \emph{TESS} have been shifted. The fitted residuals are displayed at the bottom of each panel. The complete figure in the online version of this article.}
		\label{Fig 1}
	\end{figure*}

	The phased light curves of 48 EBs and 5 pulsating variables are shown in Figure \ref{Fig 1}. The basic information of the 53 variable stars and the 10 reference stars are shown in Table 1.

	\subsubsection{\emph{TESS} observations}
	
	The \emph{Transiting Exoplanet Survey Satellite (TESS)} is an all-sky survey that observes stars brighter than $T_{\emph{mag}}$$\approx$16, with a photometric precision from 60 ppm to 3 percent, enabling an array of exoplanet and stellar astrophysics investigations \citep{a61,a62}. By cross-matching with the \emph{TESS} data, we found that 50 of 53 variable stars were observed by \emph{TESS} at 2 min cadence. We collected light curves from Mikulski Archive for Space Telescopes (MAST) but found that some targets are missing. Therefore, we carried out photometry for \emph{TESS} to obtain light curves for all targets. We used the coordinates to search and download 20 × 20 pixel \emph{TESS} full-frame image (FFI) cut-outs \citep{a64} through the \emph{lightkurve} package from the MAST archive \citep{a63}. By using the thresholding method, apertures and background masks were created to generate light curves. Low-frequency trends were removed using Scipy's Savitzky-Golay filter \citep{a65}. For LEBS-18, LEBS-21, LEBS-36, LEBS-37, LEBS-44, LEBS-45, and LEBS-48, we found that these targets were contaminated by the nearby stars due to the large pixel scale of \emph{TESS} (21 arcsec), leading to diluted amplitudes. Although we have removed the background light, the lightkurve package is currently unable to completely carry out the correction for the flux from other stars. For LEBS-37, we used the light curve provided by MAST, which completely corrects the contamination. For LEBS-18, LEBS-21, LEBS-36, LEBS-44, LEBS-45, and LEBS-48, since there are no light curves provided by MAST, we still used the light curves obtained through photometry with the lightkurve package. And for the other targets that are not contaminated by nearby stars, we used the light curves processed by lightkurve package, which were consistent with the those provided by MAST (if available). The light curves of the \emph{TESS}-band are shown with blue symbols in Figure \ref{Fig 1}.

	\begin{table*}[h!] 
		\begin{center} 
			\label{Tab 1}
			\caption{Basic information for 53 variable stars and the 10 reference stars.} 
			\setlength{\tabcolsep}{2.6mm}{    
				\resizebox{\textwidth}{10.5cm}{   
					\renewcommand\arraystretch{0.6}  
					\begin{tabular}{cccccc}  
						\hline
						\multicolumn{1}{l}{Name(This paper)} & Name(VSX website) & $\alpha_{2000}$   &$\delta_{2000}$  & \multicolumn{1}{l}{Period(d)} & \multicolumn{1}{l}{$T_{eff}$(K)} \\ \hline          
						LEBS-01               & V0663 Peg                    & 22:52:59.60 & +33:54:26.4  & 0.6244506  & 6830(±25)  \\
						LEBS-02               & ATO J343.6450+34.6956        & 22:54:35.00 & +34:41:44.3  & 1.13162    & 6776(±41)  \\
						LEBS-03               & ASASSN-V J225551.75+331017.3 & 22:55:51.75 & +33:10:17.3  & 2.39499    & 5744(±22)  \\
						LEBS-04               & V0404Peg                     & 22:56:30.90 & +33:55:12.1  & 0.419195   & 6400(±71)  \\
						LEBS-05               & V0751 Peg                    & 22:58:40.48 & +34:37:46.3  & 0.60763649 & 6765(±13)  \\
						LEBS-06               & V0557 Peg                    & 22:59:23.53 & +32:51:33.4  & 2.3845     & 4991(±93)  \\
						LEBS-07               & ASASSN-V J230252.69+342300.6 & 23:02:52.61 & +34:23:00.5  & 0.60097    & 6710(±23)  \\
						LEBS-08               & V0593 And                    & 23:03:44.07 & +36:15:22.9  & 0.372177   & 5635(±66)  \\
						LEBS-09               & ER Peg                       & 23:05:46.80 & +33:29:07.1  & 2.27467    & 7749(±62)  \\
						LEBS-10               & V0568 Peg                    & 23:08:13.01 & +33:03:03.8  & 0.247074   & 4256(±44)  \\
						LEBS-11               & ASASSN-V J231002.29+342823.4 & 23:10:02.23 & +34:28:23.2  & 0.433999   & 6523(±22)  \\
						LEBS-12               & V0747 Peg                    & 22:55:49.91 & +34:01:12.0  & 0.51431    & 7153(±54)  \\
						LEBS-13               & V0579 And                    & 22:59:11.09 & +36:21:17.8  & 0.338134   & 5538(±21)  \\
						LEBS-14               & V0478 Lac                    & 22:50:47.74 & +35:40:56.2  & 0.445665   & 6230(±98)  \\
						LEBS-15               & V0548 Peg                    & 22:51:34.19 & +34:57:53.0  & 0.38162    & 6590(±64)  \\
						LEBS-16               & ASASSN-V J225203.51+325424.4 & 22:52:03.57 & +32:54:25.4  & 0.318274   & 5877(±44)  \\
						LEBS-17(new)          & —                            & 22:52:23.27 & +33:19:21.1  & 16.94967   & 5757(±84)  \\
						LEBS-18               & ASASSN-V J225235.00+362014.1 & 22:52:35.00 & +36:20:14.2  & 1.0859832  & 7220(±54)  \\
						LEBS-19               & BD+34 4783                   & 22:52:50.63 & +35:31:58.8  & 4.118      & 6426(±64)  \\
						LEBS-20               & V0479 Lac                    & 22:52:50.69 & +35:58:56.5  & 0.34575    & 5760(±33)  \\
						LEBS-21               & ASASSN-V J225325.14+354532.5 & 22:53:25.24 & +35:45:32.8  & 0.354248   & 5380(±100) \\
						LEBS-22               & V0664Peg                     & 22:53:59.34 & +33:33:46.8  & 0.40037    & 6770(±76)  \\
						LEBS-23               & ASASSN-V J225402.95+324527.6 & 22:54:02.95 & +32:45:27.6  & 0.267356   & 4520(±81)  \\
						LEBS-24               & ASASSN-V J225412.47+335322.7 & 22:54:12.50 & +33:53:24.1  & 0.331146   & 5933(±19)  \\
						LEBS-25               & ASASSN-V J225440.29+362222.9 & 22:54:40.40 & +36:22:22.3  & 0.436834   & 6840(±73)  \\
						LEBS-26               & VW Peg                       & 22:56:23.58 & +33:13:43.8  & 21.071749  & 5630(±35)  \\
						LEBS-27               & ASASSN-V J225650.69+345101.6 & 22:56:50.70 & +34:51:01.9  & 0.33193    & 5930(±33)  \\
						LEBS-28               & V0749 Peg                    & 22:57:19.54 & +34:27:58.2  & 0.4167682  & 6657(±47)  \\
						LEBS-29               & ASASSN-V J225721.72+350704.5 & 22:57:21.87 & +35:07:03.1  & 5.1010937  & 6707(±66)  \\
						LEBS-30               & ASASSN-V J225822.44+341230.9 & 22:58:22.46 & +34:12:31.9  & 1.62663    & 7340(±137) \\
						LEBS-31               & ASASSN-V J225837.15+333030.0 & 22:58:37.15 & +33:30:30.0  & 0.255738   & 5543(±28)  \\
						LEBS-32(new)          & —                            & 22:59:12.60 & +33:30:15.0  & 1.00608444 & 4990(±164) \\
						LEBS-33               & ASASSN-V J230021.29+333532.7 & 23:00:21.24 & +33:35:32.0  & 0.53253    & 6417(±54)  \\
						LEBS-34               & V0667 Peg                    & 23:03:06.37 & +33:51:51.2  & 0.3933915  & 6303(±40)  \\
						LEBS-35               & ASASSN-V J230313.97+344622.3 & 23:03:13.83 & +34:46:21.9  & 0.4228617  & 4043(±41)  \\
						LEBS-36               & ASASSN-V J230358.08+335926.6 & 23:03:57.99 & +33:59:27.4  & 1.86598    & 7330(±93)  \\
						LEBS-37               & ASASSN-V J230416.79+360511.9 & 23:04:16.79 & +36:05:11.9  & 3.52444    & 6723(±97)  \\
						LEBS-38               & ASASSN-V J230437.49+344949.5 & 23:04:37.38 & +34:49:49.7  & 0.2901     & 5273(±50)  \\
						LEBS-39               & ASASSN-V J230506.67+325332.8 & 23:05:06.57 & +32:53:32.8  & 0.390655   & 6187(±106) \\
						LEBS-40               & ASASSN-V J230607.98+341038.9 & 23:06:08.01 & +34:10:39.0  & 0.376725   & 6097(±68)  \\
						LEBS-41               & V0669 Peg                    & 23:06:28.11 & +33:40:09.6  & 0.3970365  & 5843(±60)  \\
						LEBS-42               & ASASSN-V J230635.23+325632.1 & 23:06:35.16 & +32:56:32.0  & 0.2669558  & 5060(±109) \\
						LEBS-43               & ASASSN-V J230722.65+360358.1 & 23:07:22.67 & +36:03:58.1  & 0.2495618  & 4933(±136) \\
						LEBS-44               & ASASSN-V J230819.40+350853.0 & 23:08:19.42 & +35:08:53.2  & 0.874223   & 4830(±10)  \\
						LEBS-45               & ASASSN-V J230835.43+354143.9 & 23:08:35.33 & +35:41:43.6  & 0.299564   & 5100(±10)  \\
						LEBS-46               & CSS\_J230837.7+333905        & 23:08:37.76 & +33:39:05.5  & 0.389588   & 6400(±110) \\
						LEBS-47               & CSS\_J230838.3+334753        & 23:08:38.34 & +33:47:53.0  & 0.8839762  & 6601(±33)  \\
						LEBS-48               & ASASSN-V J230923.71+361629.8 & 23:09:23.71 & +36:16:30.1  & 0.314522   & 5377(±50)  \\
						Pulsatingvariablestar & CSS\_J225208.2+355354        & 22:52:08.20 & +35:53:54.8  & 0.7324429  & —          \\
						Pulsatingvariablestar & CSS\_J225940.9+325805        & 22:59:40.90 & +32:58:05.1  & 0.6087498  & —          \\
						Pulsatingvariablestar & CSS\_J230156.0+360059        & 23:01:56.00 & +36:00:59.7  & 0.5083755  & —          \\
						Pulsatingvariablestar & V0564 Peg                    & 23:03:10.27 & +34:25:07.1  & 0.318785   & —          \\
						Pulsatingvariablestar & CSS\_J230536.3+344154        & 23:05:36.42 & +34:41:54.1  & 0.198771   & —          \\
						Comparison star       & —                            & 23:06:49.00 & +36:09:01.3  & —          & —          \\
						Check star            & —                            & 22:57:21.57 & +35:22:00.64 & —          & —          \\
						Check star            & —                            & 22:57:23.03 & +35:25:00.00 & —          & —          \\
						Check star            & —                            & 22:58:17.74 & +33:30:27.25 & —          & —          \\
						Check star            & —                            & 22:58:26.06 & +33:29:18.85 & —          & —          \\
						Check star            & —                            & 23:04:53.39 & +36:08:37.18 & —          & —          \\
						Check star            & —                            & 23:05:49.00 & +33:00:46.7  & —          & —          \\
						Check star            & —                            & 23:05:50.03 & +33:05:03.29 & —          & —          \\
						Check star            & —                            & 23:06:49.07 & +36:09:01.26 & —          & —          \\
						Check star            & —                            & 23:06:57.53 & +36:09:55.56 & —          & —          \\
						Check star            & —                            & 23:06:58.00 & +36:09:55.5  & —          & —          \\
						Check star            & —                            & 23:06:58.86 & +34:52:05.79 & —          & —            \\ \hline
			\end{tabular}}}
			\begin{tablenotes}    
				\footnotesize           
				\item[1] "Name(This paper)" represents the name used in this paper. "Name(VSX website)" represents the name on the VSX website. "$\alpha_{2000}$" and "$\delta_{2000}$" represent the right ascension and declination of EBs, respectively. "Period(d)" represents the orbital period for EBs and the pulsation period for pulsating variable stars, the periods for LEBS-02, LEBS-17, and LEBS-32 are derived from the Lomb-Scargle method, while the rest are obtained through queries on the VSX website. "$T_{eff}(K)$" represents the effective temperature of the EBs, those for LEBS-01 to LEBS-13 are from the LAMOST spectra, and for LEBS-14 to LEBS-48 are determined using the de-reddened color index.
			\end{tablenotes}     
		\end{center}
	\end{table*}

	\subsection{Spectroscopy}
	\subsubsection{LAMOST Spectral Observations}
	
	LAMOST is a 4m reflecting Schmidt optical telescope at Xinglong Observatory, with a 5° field of view \citep{a12lamost,a13lamost}. The focal plane can accommodate up to 4000 optical fibers connected to 16 identical multi-objective optical spectrometers. LAMOST Low-Resolution Spectroscopic Survey (LRS) has a resolution of R$\sim$1800, and the wavelength coverage is approximately 3700-9000Å. The Medium-Resolution Spectroscopic Survey (MRS) has a resolution of R$\sim$7500. In the MRS mode, the wavelength coverage of the blue arm is 4950-5350 Å, and the coverage of the red arm is 6300-6800 Å \citep{a14lamost}. By cross-matching 48 EBs with LAMOST DR9, we obtained high-quality (S/N$\textgreater$10) medium-resolution spectra of 13 EBs (LEBS-01 $\sim$ LEBS-13),  which were used for subsequent analysis.

	\subsubsection{Radial velocity analysis}
	
\begin{figure*}[ht!]
		\centering
		\includegraphics[width=5.5cm]{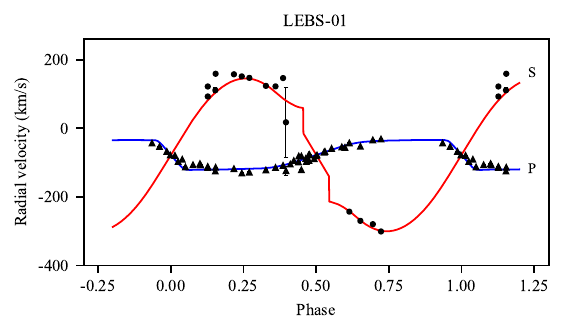}
		\includegraphics[width=5.5cm]{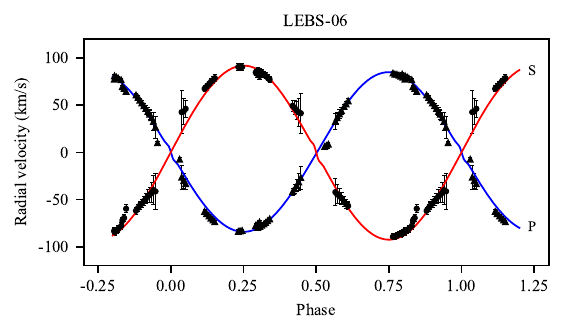}
		\includegraphics[width=5.5cm]{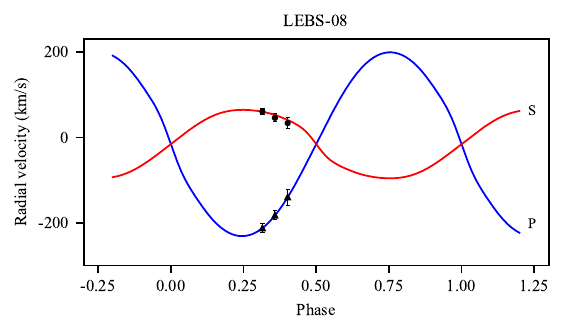}
		\caption{Theoretical and observational radial velocity curves of 13 targets. The blue lines show the theoretical curves of the primary components, and the red lines show the theoretical curves of the secondary components. The complete figure in the online version of this article. }
		\label{Fig 2}
	\end{figure*}

	To avoid the influence of possible H$\alpha$ emission line, only blue arm spectra were used to measure the radial velocities. Synthetic PHOENIX spectra \citep{a17rv} with reduced resolution to the same resolution as the LAMOST spectra (R $\sim$ 7500) were used as template spectra depending on the temperature (the calculation of temperatures of the two components is described in Section 3) of each EB. We used the cross-correlation function (CCF) method \citep{a15rv,a16rv} to measure the radial velocities. The GaussPy+ package \citep{a69} was used to determine the peak position of the CCF curves. We used the method proposed by \cite{a18rv} for zero point correction of radial velocities. Rvfit code \citep{a68} was used to obtain the phased radial velocity curves, which are shown in Figure \ref{Fig 2}. The radial velocities of the 13 targets are listed in Table 2. As seen in Figure \ref{Fig 2}, the radial velocities of LEBS-12 and LEBS-13 show more scatter due to a limitation of the data, which were not used for orbital parameter study, and the results of the other 11 targets were used for subsequent orbital parameter analysis. For LEBS-03 and LEBS-08, radial velocity measurements were obtained at only three phases, which makes it difficult to obtain a unique solution. But in Section 3, by combing the radial velocity curves with the photometric light curves, we can obtain reliable orbital parameters, as the light curves provide crucial supplementary information. The large dispersion of LEBS-06 and LEBS-07 at phases 0 and 0.5 is due to poor data quality of the spectra. All 13 systems are all double-lined spectroscopic binaries, including LEBS-09, even though the primary component's luminosity accounts for more than 90\% of the total (please refer to Section 3 for details), its CCF curve does indeed show double peaks near the phases of 0.25 and 0.75. Moreover, based on the formula $m_{2} - m_{total} = -2.5lg\frac{L_{2}}{L_{total}} ,m_{total}=10.744$ (G-band) \citep{a144}, $\frac{L_{2}}{L_{total}} = 0.093$ (\emph{TESS}-band, which is very close to the G-band), we calculated the magnitude of the secondary component, $m_{2}$=13.352(±0.123), which is within the observation limit of LAMOST (m$\approx$15, G-band). Therefore, LEBS-09 is a double-lined spectroscopic binary.

	\begin{table*}[h!] 
		\centering
		\huge  
		\label{Tab 2}
		\caption{The radial velocity values of 13 EBs.} 
		\setlength{\tabcolsep}{1.5cm}{    
			\resizebox{\textwidth}{2.5cm}{   
				\renewcommand\arraystretch{1.0}  
				
				\begin{tabular}{ccccccccc}  
					
					\hline
					\multicolumn{1}{l}{LEBS-01} &HJD  &phase & RV-P(km/s)  &Error(km/s) & RV-S(km/s) &Error(km/s)  \\ \hline  
					& 2458408.03318 & 0.100 & -103.67     & 0.90  & -           & -       \\
					& 2458408.04951 & 0.127 & -114.60     & 1.07  & 92.76       & 4.29   \\
					& 2458408.06645 & 0.153 & -125.41     & 1.57  & 110.90      & 7.26    \\
					& 2458408.10554 & 0.217 & -118.35     & 0.72  & 157.43      & 1.68    \\
					& 2458408.12299 & 0.244 & -131.21     & 0.99  & 151.36      & 2.52  \\
					& 2458408.13932 & 0.270 & -128.59     & 1.39  & 146.98      & 3.25   \\
					& 2458410.06862 & 0.360 & -115.08     & 0.88  & 122.21      & 2.28     \\
					& 2458410.08485 & 0.386 & -108.95     & 0.69  & 146.48      & 3.23    \\
					& 2458416.09653 & 0.014 & -79.78      & 0.86  & -           & -        \\
					& 2458416.11276 & 0.039 & -90.99      & 0.80  & -           & -        \\
					& 2458446.93441 & 0.396 & -124.60     & 12.84 & 17.55       & 101.74   \\
					& 2458447.95055 & 0.025 & -97.44      & 0.70  & -           & -      
					\\ \hline
		\end{tabular}}}
		\begin{tablenotes}    
			\footnotesize           
			\item[1] "HJD" represents the observation time. “Phase” represents the phase used in this work, which is determined based on the ephemeris calculated in our work. "RV-P(km/s)" and "RV-S(km/s)" represent the radial velocity of the primary and secondary components, respectively. 
			\item[2] This table is available in its entirety in machine-readable form through China VO (https://nadc.china-vo.org/res/r101429/). 
		\end{tablenotes}     
	\end{table*}
	
	\section{Orbital Parameters}
	
	The 2015 version of W-D code \citep{a21,a22,a23} was used to determine the orbital parameters of 48 EBs. 
	When setting parameters, the calculation rules for the temperatures of the primary and secondary components are as follows. Firstly, the temperature of the primary components are estimated: (1) For the targets with LAMOST spectra, we used the LASP temperatures \citep{a19lasp} given by LAMOST spectra. Considering the wavelength coverage of the low-resolution spectra was wider than that of the medium-resolution spectra, we first used the temperatures given by the low-resolution spectra. For targets with only medium-resolution spectra, we used the temperatures from the medium-resolution spectra. The average value was taken as the final temperature if the target has multiple temperatures.(2) For the targets without LAMOST spectra, we used the dereddening color index, considering 3D extinction \citep{a24}. Based on the magnitudes of B, V, g, r bands from AAVSO Photometric All Sky Survey \citep{a25} and J, K bands from 2MASS \citep{a26}, we calculated B-V, g-r, J-K. According to the relationship between color index and temperature \citep{a100}, three temperatures were obtained, and we adopted the average temperature as the final temperature. However, the temperatures estimated by the LASP or color index are actually the average values of the two components. Therefore, these values were initially set as the temperature of the primaries; when we have obtained the temperature ratio, $T_{2i}/T_{1i}$, and the radius ratio, $k = r_{2}/r_{1}$, through the convergent solution of the W-D code, the more accurate temperature of each component ($T_{1c}$, $T_{2c}$) can be calculated using the following equations \citep{a116,a117}:
	\begin{equation}\label{eq1}
		T_{1c} = (\frac{(1+k^{2})T_{1i}^{4})}{1+k^{2}(\frac{T_{2i}}{T_{1i}})^{4}})^{0.25},  
	\end{equation}
	\begin{equation}\label{eq2}
		T_{2c}=T_{1c}\frac{T_{2i}}{T_{1i}}.   
	\end{equation}
	Particularly, the temperatures obtained through the LAMOST spectra and determined using the color index are not exactly corresponding to quadrature (which is assumed in the above equations), but slightly differ from the temperatures corresponding to quadrature. Although we cannot obtain exact temperatures that correspond to quadrature, we try to use these temperatures from LAMOST spectra or color index as initial values, and then calculate the more accurate temperatures of the two compenents based on Equation \ref{eq1} and \ref{eq2}.
	Then, $T_{1c}$ and $T_{2c}$ were set as initial values, with the primary temperature set as a fixed parameter and the secondary temperature as an adjustable one. Recalculations were made using the W-D code, the convergent solutions obtained are listed in Table 3.
	
	The gravity darkening and bolometric albedo coefficients were set to g = 0.32 and A = 0.5 for a convective atmosphere ($T_{eff}$ $\textless$ 7200 K), and g = 1 and A = 1 for a radiative atmosphere ($T_{eff}$ $\textgreater$ 7200 K) \citep{a27,a104,a105}. The limb darkening and the bolometric coefficients of the two components were determined according to the table of \cite{a28}. We set the mode based on the different characteristics of each eclipsing binary: Mode 2 (for detached system), Mode 3 (for overcontact system), Mode 4 (for semi-detached system with the more massive one filling its Rochelobe), and Mode 5 (for semi-detached system with the less massive star filling its Roche lobe).
	
	The mass ratio (q) plays an important role in studying the Roche-geometry and physical properties of EBs  (\citeauthor{a101} \citeyear{a101}; \citeauthor {a29} \citeyear{a29}). For the 11 targets with both light curves and radial velocity curves, we could obtain accurate mass ratios and determine absolute physical parameters by W-D code. The initial mass ratios obtained from the fitting of rvfit code. For the 37 targets with only light curves, the rules for the initial mass ratios were as follows: (1) For the detached binaries, the mass ratios are insensitive to the light curves \citep{a30,a31}, so we adopted the method provided by \cite{a32} to estimate their initial mass ratios. Assuming that the components of the EBs are main-sequence stars and the primary temperature $T_{1}^{i} =T_{eff}$, the temperature of the secondary component can be estimated as $T_{2}^{i} = T_{1}^{i}(\frac{\mathrm{d_{2}}}{\mathrm{d_{1}}} )^{0.25}$, where $d_{1}$ and $d_{2}$ are the depths of the eclipses in relative flux. Then, the mass ratios of detached binaries are $q = (\frac{\mathrm{T_{2}^{i} }}{\mathrm{T_{1}^{i}}} )^{1.6} $. (2) For semi-detached and contact binaries, the initial mass ratios were determined by the q-search method. We carried out calculations over a series of q-values, ranging from 0.1 to 10.0 with a step of 0.1. For each assumed q-value, a convergent solution and mean residuals were obtained. We treated the q-value corresponding to the smallest mean residuals as the initial mass ratio of the system. The q-search method is widely used to determine mass ratios of binaries when radial velocity curves are not required \citep{a33,a34,a35,a112}. 
	
	\begin{table*}[ht!]
		\centering
		\label{Tab 3}
		\caption{Orbital parameters of 48 EBs from W-D solutions.} 
		\begin{tabular}{cccccc}
			\hline
			Parameters     & LEBS-01         & LEBS-02         & LEBS-03         & LEBS-04         & LEBS-05         \\\hline
			Mode               & 3               & 3               & 2               & 3               & 3               \\
			$T_{1}$(K)         & 6972(±30)       & 6862(±44)       & 5813(±32)       & 6426(±84)       & 6884(±16)       \\
			$T_{2}$(K)         & 6183(±57)       & 6604(±123)      & 5688(±57)       & 6292(±161)      & 6193(±32)       \\
			i($^{\circ}$)      & 89.058(±0.158)  & 36.541(±0.239)  & 89.248(±0.090)  & 61.645(±0.088)  & 69.527(±0.083)  \\
			q($M_{2}$/$M_{1}$) & 0.220(±0.001)   & 0.632(±0.004)   & 2.210(±0.020)   & 0.214(±0.002)   & 0.190(±0.001)   \\
			$\Omega_{1}$       & 2.242(±0.003)   & 3.071(±0.015)   & 8.843(±0.026)   & 2.211(±0.006)   & 2.180(±0.002)   \\
			$\Omega_{2}$       & 2.242(±0.003)   & 3.071(±0.015)   & 13.250(±0.105)  & 2.211(±0.006)   & 2.180(±0.002)   \\
			$\Omega_{in}$      & 2.282           & 3.132           & 5.547           & 2.267           & 2.207           \\
			$\Omega_{out}$     & 2.143           & 2.762           & 4.943           & 2.131           & 2.085           \\
			f($f_{1}$)(\%)     & 29.2(±2.1)      & 11.0(±6.4)      & 62.7(±0.2)      & 41.0(±4.4)      & 22.2(±1.7)      \\
			$f_{2}$(\%)        & —               & —               & 41.9(±0.3)      & —               & —               \\
			$r_{1}$            & 0.527(±0.001)   & 0.431(±0.003)   & 0.152(±0.000)   & 0.536(±0.001)   & 0.534(±0.000)   \\
			$r_{2}$            & 0.271(±0.003)   & 0.320(±0.008)   & 0.172(±0.001)   & 0.269(±0.007)   & 0.263(±0.003)   \\
			$L_{1r}$/$L_{r}$   & 0.856(±0.001)   & 0.640(±0.005)   & 0.458(±0.001)   & 0.806(±0.002)   & 0.867(±0.001)   \\
			$L_{1i}$/$L_{i}$   & 0.846(±0.001)   & 0.633(±0.005)   & 0.455(±0.001)   & 0.804(±0.002)   & 0.859(±0.001)   \\
			$L_{1t}$/$L_{t}$   & 0.847(±0.001)   & 0.634(±0.005)   & 0.455(±0.001)   & 0.804(±0.002)   & 0.860(±0.001)   \\
			Latitude (star1)   & 0.571           & 1.598           & —               & —               & —               \\
			Longitude (star1)  & 0.478           & 2.389           & —               & —               & —               \\
			Radius (star1)     & 0.257           & 0.330           & —               & —               & —               \\
			$T_{s}$ (star1)    & 0.880           & 1.202           & —               & —               & —               \\
			$V_{\gamma}$(km/s) & -78.570(±1.293) & -11.990(±2.300) & -59.700(±5.199) & -16.810(±2.064) & -57.810(±0.667) \\
			a(R$_{\odot}$)     & 3.524(±0.103)   & 7.399(±0.417)   & 9.306(±0.444)   & 2.509(±0.077)   & 3.264(±0.042)   \\\hline
		\end{tabular}
		\begin{tablenotes}    
			\footnotesize           
			\item[1] "Mode" represents the configuration mode of EBs, where "2" denotes a detached system, "3" denotes an overcontact system, "4" denotes a semi-detached system with the more massive star filling its Roche Lobe, and "5" denotes a semi-detached system with the less massive star filling its Roche Lobe. "$T_{1}$(K)" and "$T_{2}$(K)" represent the temperature of the primary and secondary components, respectively. "i($^{\circ}$)" represents the orbital inclination. "q($M_{2}$/$M_{1}$)" represents the mass ratio. "$\Omega_{1}$" and "$\Omega_{2}$" represent the dimensionless potential energy of the primary and secondary components, respectively. "$\Omega_{in}$" and "$\Omega_{out}$" represent the potential energy of the inner critical surface and the outer critical surface, respectively. "f(\%)" represents the contact degree for contact binaries. "$f_{1}$(\%)" and "$f_{2}$(\%)" represent the fill-out factor of the primary and secondary components for detached and semi-detached binaries. "$r_{1}$" and "$r_{2}$" represent the relative radius of the primary and secondary components, which are their radii relative to the semi-major axis, respectively. "$L_{1r}$/$L_{r}$", "$L_{1i}$/$L_{i}$" and "$L_{1t}$/$L_{t}$" represent the share of the primary component's luminosity in the Sloan r', Sloan i' and \emph{TESS} bands, respectively. "$L_{3r}$/$L_{r}$, "$L_{3i}$/$L_{i}$ and "$L_{3t}$/$L_{t}$" represent the share of the third light's luminosity in the Sloan r', Sloan i' and \emph{TESS} bands, respectively. "Latitude", "Longitude", "Radius" and "$T_{s}$" represent the latitude, longitude, angular radius and temperature factor of the starspots, respectively. "$V_{\gamma}$ (km/s)" represents the radial velocity of the system. "a(R$_{\odot}$)" represents the semi-major axis. "e" represents eccentricity. "$\omega$" represents the periastron angle. "-" indicates the corresponding parameter can't be obtained for the EBs. 
			
			\item[2] This table is available in its entirety in machine-readable form through China VO (https://nadc.china-vo.org/res/r101429/). 
		\end{tablenotes}     
	\end{table*}
	
	In the process of solving the orbital parameters, there are several special types of targets, for which the analysis was as follows: (1) For LEBS-18, LEBS-21, LEBS-36, LEBS-37, LEBS-44, LEBS-45, LEBS-48, the amplitudes of the \emph{TESS}-band light curves were diluted due to the contamination of nearby stars resulting from the large pixel scale (21 arcsec) of \emph{TESS}. The third light would make the photometric solution complex and uncertain \citep{a36}. We first estimated the physical parameters through the Sloan r' and i' bands. Then, we used these physical parameters and set the third light to be adjustable for the \emph{TESS} band to determine the third light. Next, we set the third light of the \emph{TESS} band to non-adjustable, and the final solutions were determined by analyzing the light curves of the three bands. (2) For LEBS-01, LEBS-02, LEBS-06, LEBS-08, LEBS-10, LEBS-14, LEBS-20, LEBS-22, and LEBS-40, the two maxima of the light curves showed asymmetric characteristics. This phenomenon is the so-called "O'Connell effect" \citep{a37}. We added starspots on one or both components to address that. (3) For the three detached binaries, LEBS-17, LEBS-26, and LEBS-29, the light curves showed that the orbits have apparent eccentricity. Eccentricity e and periastron angle $\omega$ were calculated by the following equations \citep{a38},
	\begin{equation}\label{eq3}
		e_{0}\cos\omega_{0} = \frac{\pi}{2} [(\varphi_{2} -\varphi_{1})-0.5],
	\end{equation}
	\begin{equation}\label{eq4}
		e_{0}\sin\omega_{0} = \frac{w_{2}-w_{1}}{w_{2} +w_{1} },
	\end{equation}
	where $\varphi_{1,2}$ are the phase of primary and secondary minima, and $w_{1,2}$ are the widths of them.
	
	The orbital parameters from the W-D solutions are listed in Table 3. The orbital parameter solutions for total eclipse contact and semi-detached binaries, as well as for systems with radial velocity curves, are uniquely determined (including LEBS-01 to LEBS-11, LEBS-21, LEBS-22, LEBS-28, LEBS-34, LEBS-41, and LEBS-48). The solutions for other systems are not unique; the solutions presented in Table 3 are the optimal solutions (including LEBS-12 to LEBS-20, LEBS-23 to LEBS-27, LEBS-29 to LEBS-33, LEBS-35 to LEBS-40, and LEBS-42 to LEBS-47). The uncertainties in the orbital parameters are fitting errors, which are generally underestimated \citep{a134}. Some errors are reported as 0.000, which are attributed to rounding effects. The theoretical light curves (black lines) and the fitting residuals are shown in Figure \ref{Fig 1}. The theoretical radial velocity curves (blue and red lines represent primary and secondary components, respectively) considering proximity effect are shown in Figure \ref{Fig 2}.

	\section{Orbital Period Investigations}
	
	Studying orbital period variations promotes the research of dynamic interactions of binary systems, searching for third bodies, and material transfer between two components of binary systems \citep{a39,a40,a41}. For the 26 targets with sufficient eclipsing times, we calculated the O-C (observed eclipsing minimum minus calculated eclipsing minimum) values to analyze their orbital period variations. We recorded the eclipsing minima from the O–C gateway\footnote{\url{http://var2.astro.cz/ocgate/}} and previous studies, collected many photometric survey light curves and then used the K–W method \citep{a42} to calculate the minima, including the following surveys: the All-Sky Automated Survey for SuperNovae (ASAS-SN, \citeauthor{a43} \citeyear{a43}; \citeauthor{a44} \citeyear{a44}; \citeauthor{a45} \citeyear{a45}), the Zwicky Transient Facility (ZTF) survey (\citeauthor{a46} \citeyear{a46}; \citeauthor{a47} \citeyear{a47}), Wide Angle Search for Planets (WASP, \citeauthor{a48} \citeyear{a48}). We also used data from the American Association of Variable Star Observers (AAVSO,\citeauthor{a49} \citeyear{a49}). Since the ASAS-SN and ZTF observations were discrete, we used the Li method \citep{a50} to group and fold the data into one period and then calculate the minima. We used the online tool\footnote{\url{https://astroutils.astronomy.osu.edu/time/hjd2bjd.html}} to convert HJD to BJD for all surveys uniformly. The time standard of the BJD values is TDB. All minima, errors, and their sources are listed in Table 4.

	\begin{table*}[ht!] 
		\huge  
		\label{Tab 4}
		\caption{ The eclipsing times of 26 EBs.} 
		\setlength{\tabcolsep}{1.85cm}{    
			\resizebox{\textwidth}{2.65cm}{   
				\setlength{\tabcolsep}{4pt}
				\renewcommand\arraystretch{1.1}  
				\begin{tabular}{ccccccccccccc}  
					\hline
					\multicolumn{1}{l}{LEBS-01} & BJD           & Error   & Epoch    & O-C      & Residuals & Source & BJD           & Error   & Epoch    & O-C      & Residuals & Source\\ \hline 
					& 2453226.57889 & 0.00188 & -10567.5 & -0.01869 & -0.00376  & (1)    & 2454022.44175 & 0.00242 & -9293   & -0.01812 & -0.00496  & (1)    \\
					& 2453227.52106 & 0.00168 & -10566   & -0.01320 & 0.00173   & (1)    & 2454057.41571 & 0.00189 & -9237   & -0.01340 & -0.00031  & (1)    \\
					& 2453232.51913 & 0.00207 & -10558   & -0.01074 & 0.00419   & (1)    & 2454295.64458 & 0.00206 & -8855.5 & -0.01243 & 0.00013   & (1)    \\
					& 2453237.51414 & 0.00196 & -10550   & -0.01133 & 0.00358   & (1)    & 2454295.64476 & 0.00205 & -8855.5 & -0.01225 & 0.00031   & (1)    \\
					& 2453242.50841 & 0.00195 & -10542   & -0.01266 & 0.00224   & (1)    & 2454319.68582 & 0.00147 & -8817   & -0.01254 & -0.00004  & (1)    \\
					& 2453256.55131 & 0.00243 & -10519.5 & -0.01990 & -0.00503  & (1)    & 2454326.55438 & 0.00113 & -8806   & -0.01294 & -0.00045  & (1)    \\
					& 2453262.48550 & 0.00131 & -10510   & -0.01799 & -0.00314  & (1)    & 2454329.67743 & 0.00102 & -8801   & -0.01214 & 0.00034   & (1)    \\
					& 2453265.61234 & 0.00168 & -10505   & -0.01341 & 0.00144   & (1)    & 2454330.61769 & 0.00214 & -8799.5 & -0.00856 & 0.00392   & (1)    \\
					& 2453271.54323 & 0.00228 & -10495.5 & -0.01480 & 0.00004   & (1)    & 2454331.55086 & 0.00125 & -8798   & -0.01206 & 0.00042   & (1)    \\
					& 2453950.63662 & 0.00112 & -9408    & -0.01144 & 0.00189   & (1)    & 2454334.67625 & 0.00105 & -8793   & -0.00892 & 0.00355   & (1)    \\
					& 2453970.61679 & 0.00133 & -9376    & -0.01368 & -0.00040  & (1)    & 2454350.59320 & 0.00142 & -8767.5 & -0.01546 & -0.00303  & (1)    \\
					& 2453980.60857 & 0.00122 & -9360    & -0.01311 & 0.00014   & (1)    & 2454351.53394 & 0.00120 & -8766   & -0.01140 & 0.00103   & (1)    \\
					& 2453990.60279 & 0.00241 & -9344    & -0.01010 & 0.00313   & (1)    & 2454355.58875 & 0.00176 & -8759.5 & -0.01552 & -0.00309  & (1)    \\
					& 2453997.46588 & 0.00111 & -9333    & -0.01597 & -0.00275  & (1)    & 2454360.58175 & 0.00221 & -8751.5 & -0.01812 & -0.00571  & (1)    \\
					& 2454005.58144 & 0.00218 & -9320    & -0.01827 & -0.00506  & (1)    & 2454361.52335 & 0.00119 & -8750   & -0.01320 & -0.00079  & (1)
					\\ \hline
		\end{tabular}}}
		\begin{tablenotes}    
			\footnotesize           
			\item[1] "BJD" represents the observed eclipsing minima times. "Epoch" represents the cycle number. "O-C" and "Residuals" represent the observed minima minus calculated minima and the corresponding fitting residuals, respectively. "Source" represents the source of the observed minima. (1) WASP; (2)AAVSO; (3)ASAS-SN; (4)O-C Gateway; (5)Other paper; (6)ZTF; (7)This paper; (8)\emph{TESS}       
			\item[2] This table is available in its entirety in machine-readable form through China VO (https://nadc.china-vo.org/res/r101429/). 
		\end{tablenotes}       
	\end{table*}

\begin{figure*}[htp!]
		\centering
		\includegraphics[width=7cm]{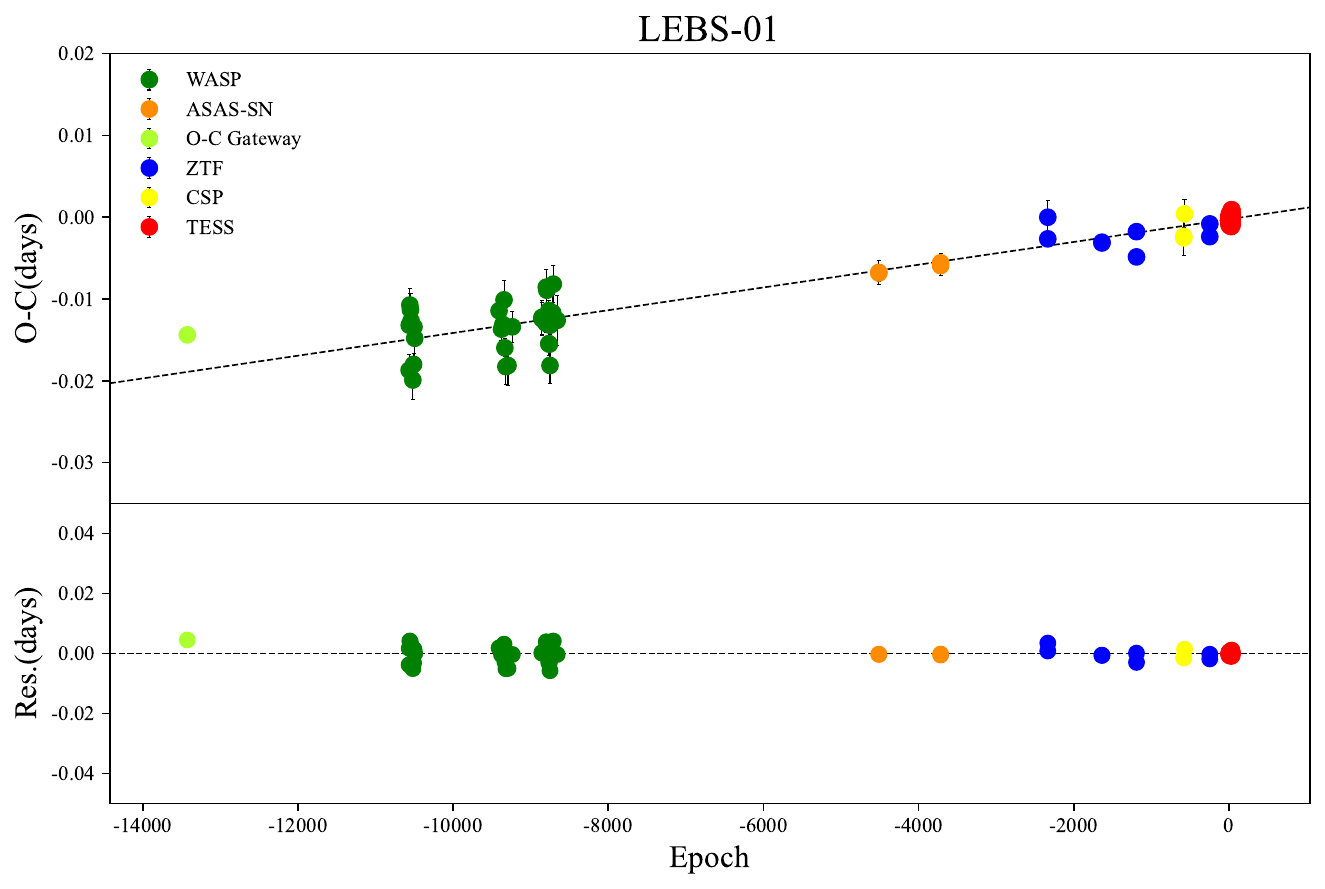}
		\hspace{1.35cm} 
		\includegraphics[width=7cm]{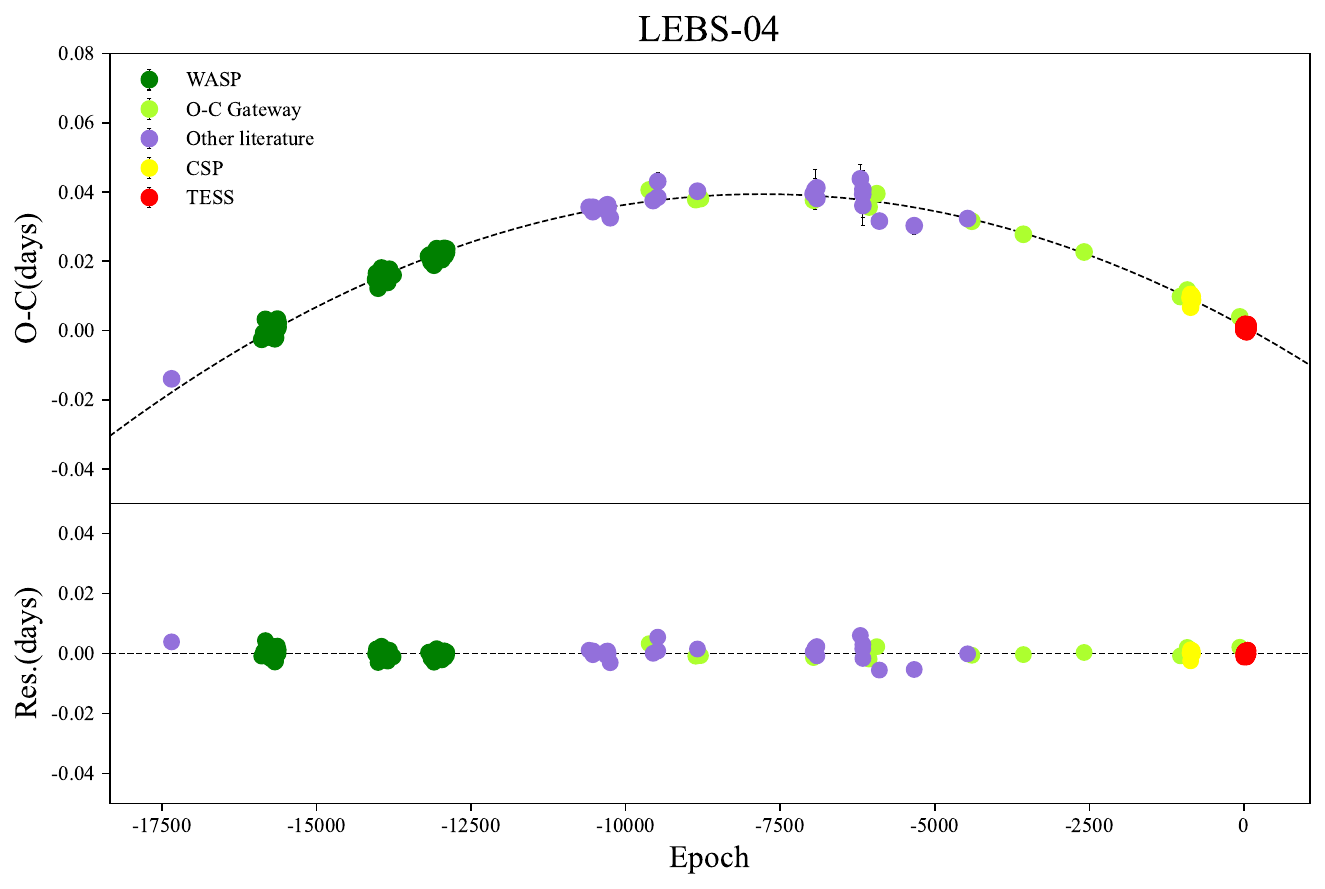}
		\includegraphics[width=7cm]{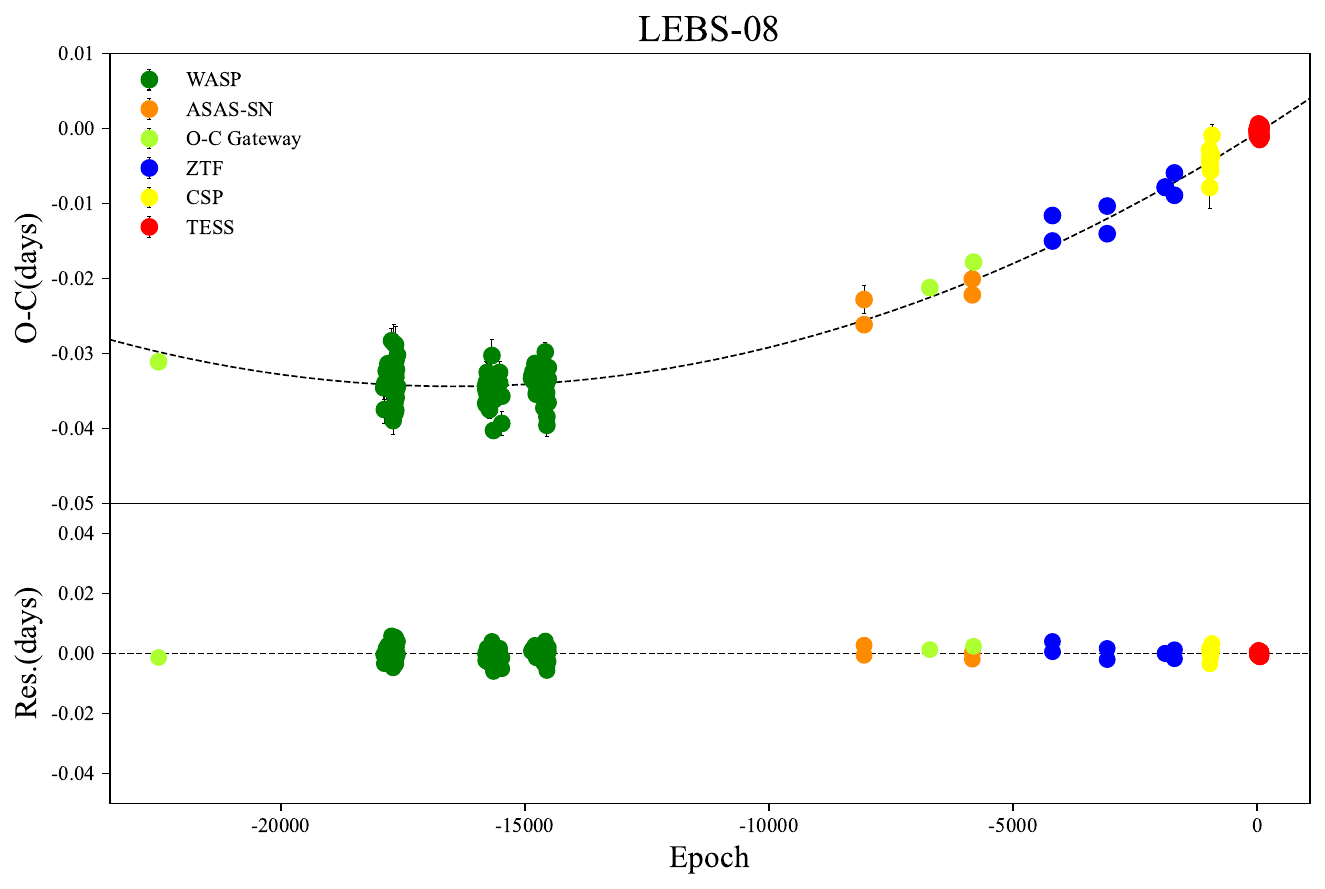}
		\hspace{1.35cm}
		\includegraphics[width=7cm]{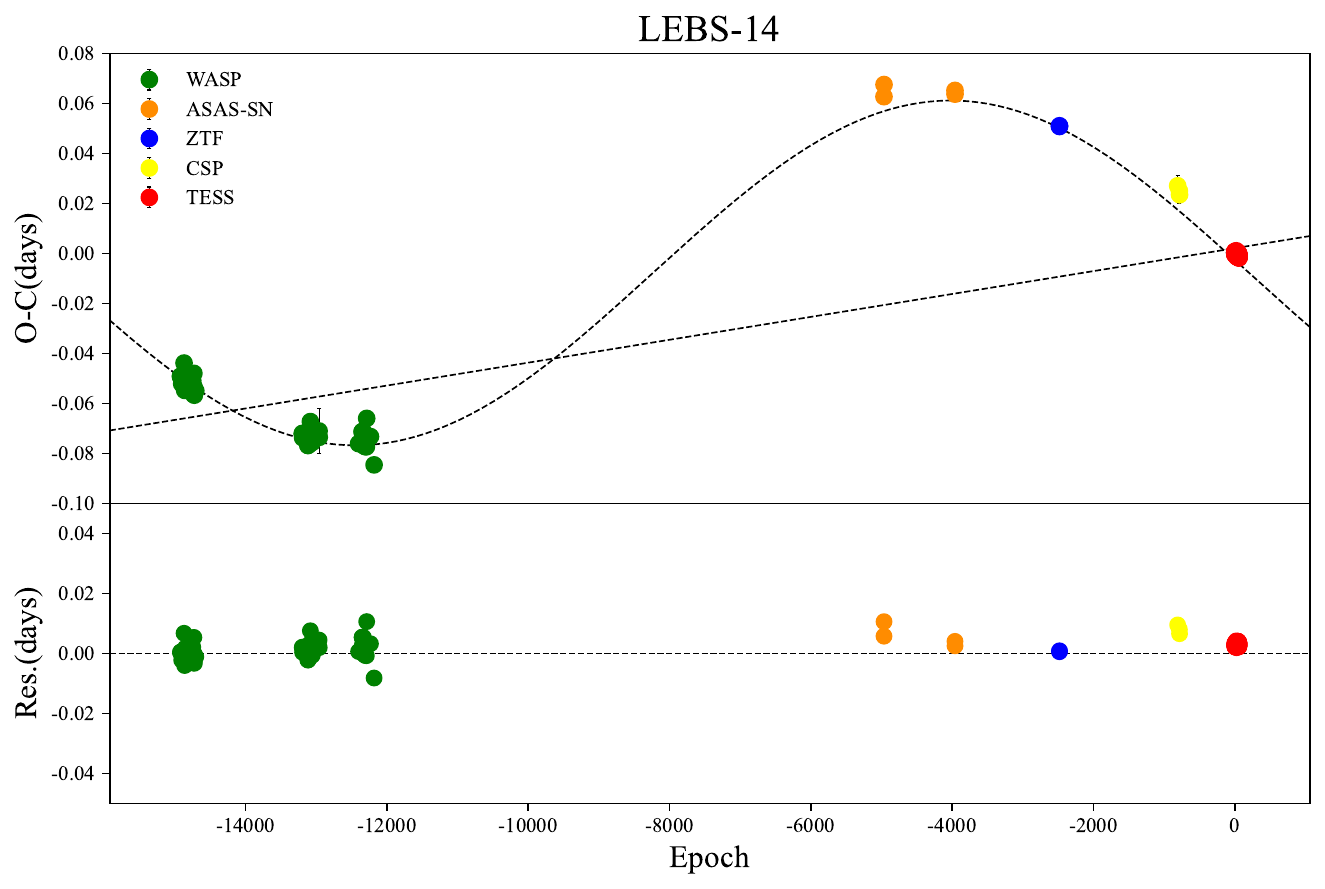}
		\caption{The O-C curves of 26 EBs. The O-C values and the fitting curves are shown at the top panels, and the residuals are at the bottom. The complete figure in the online version of this article.}
		\label{Fig 3}
	\end{figure*}

	We calculated the O-C values according to the following equation, 
	\begin{equation}\label{eq3}
		BJD = BJD_{0}  + P \times E, 
	\end{equation}
	where E represents the cycle number. The O-C values of 26 targets are plotted in Figure \ref{Fig 3}, among which the O-C curves of 16 targets exhibit a parabolic trend, 8 targets show a linear trend, and 2 targets (LEBS-14 and LEBS-41) display periodic variations. The fitting parameters of O-C curves followed Equation \ref{eq4} except for LEBS-14 and LEBS-41, 
	\begin{equation}\label{eq4}
		O-C = \Delta T_{0} + \Delta P_{0} \times E +\frac{\beta}{2}\times E^{2},
	\end{equation}
	where $\Delta T_{0}$ and $\Delta P_{0}$  represent the corrections of the initial epoch and the orbital period, $\beta$ represents the long-term changing rate of the orbital period. For LEBS-14 and LEBS-41, the O - C curves showed periodic variations, so the fitting parameters followed Equation \ref{eq5},
	\begin{equation}\label{eq5}
		O-C = \Delta T_{0} + \Delta P_{0} \times E +A\times \sin (\frac{2\pi }{P_{mod} } \times E+\varphi ),
	\end{equation}
	where A, $P_{mod}$, and $\varphi$ represent the quasi-sinusoidal change's amplitude, period, and initial phase. The fitting parameters of 26 targets were calculated as shown in Table 5.
	
	There are special explanations for some targets. (1) Due to the incorrect period, the O-C curves of LEBS-04, LEBS-10, LEBS-20, and LEBS-22 showed trends in multiple segments. We selected the most prominent trend to correct the period and recalculate the O-C values using the corrected period. This method has been addressed in many studies, such as \cite{a50}. (2) Due to the eccentric orbits, the O-C values of the primary and secondary minima of LEBS-26 and LEBS-29 could not be fitted with a single line. Both the primary and secondary minima of LEBS-29 showed a linear trend with slopes less than zero. The primary minima of LEBS-26 showed a linear trend with slopes more than zero, and the secondary components showed less than zero, which was the effect of the apsidal motion \citep{a103}.

	\section{Chromospheric Activity Investigations}
	Chromospheric activity can influence material and energy transfer in binary systems, potentially affecting their evolution \citep{a135}. Therefore, the study of chromospheric activity in EBs is of great significance for revealing the mechanisms of internal energy release in stars, understanding the structural changes during the process of stellar evolution, and verifying the theories of stellar evolution \citep{a140}. We analyzed the H$\alpha$ (6562.5Å) emission line, whose equivalent width (EW) can effectively reflect the stellar chromospheric activity level \citep{a51}. Due to the influence of the photospheric layer, it was difficult to observe H$\alpha$ emission lines directly. Therefore, the chromospheric emission lines were estimated by subtracting the photospheric contribution from the system spectra \citep{a52}. We analyzed chromospheric activity for the 13 targets with LAMOST medium-resolution spectra as follows:
	
	(1) Selected two templates and downloaded the spectra. The templates were selected from the inactive star catalog from \cite{a18rv}. The temperature difference between the two templates and the two components of the target was less than 100 K.
	
	(2) Normalized the target and template spectra and removed the cosmic rays. 
	
	(3) Generated synthetic spectra. We used the FORTRAN code STARMOD developed by \cite{a52}, which was used by many researchers (e.g. \citeauthor{a53} \citeyear{a53}; \citeauthor{a54} \citeyear{a54}; \citeauthor{a55} \citeyear{a55}). The program synthesized the spectra by treating the two nonactive template spectra as the primary and secondary components, respectively, and took into account the effects of rotation, radial velocity, and the luminosity ratio. We initially assumed synchronous rotation, the rotation speed was estimated to be used as an input value for STARMOD, and then STARMOD fits the spectra to obtain the best results.
	
	(4) Obtained the subtracted spectra by subtracting the synthesized spectra from the target spectra. Gaussian fit was performed on the H$\alpha$ emission lines of the subtraction spectra to calculate equivalent width (EW).
	
\begin{figure*}[htp]
		\centering
		\includegraphics[width=7cm]{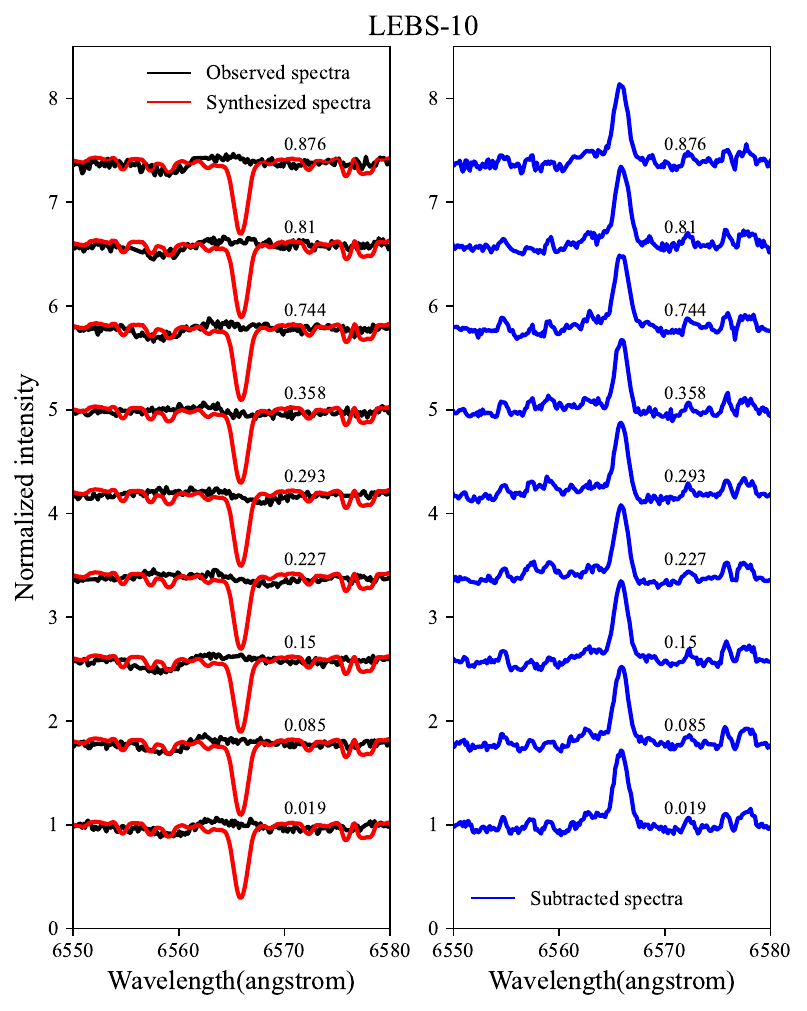}
		\includegraphics[width=7cm]{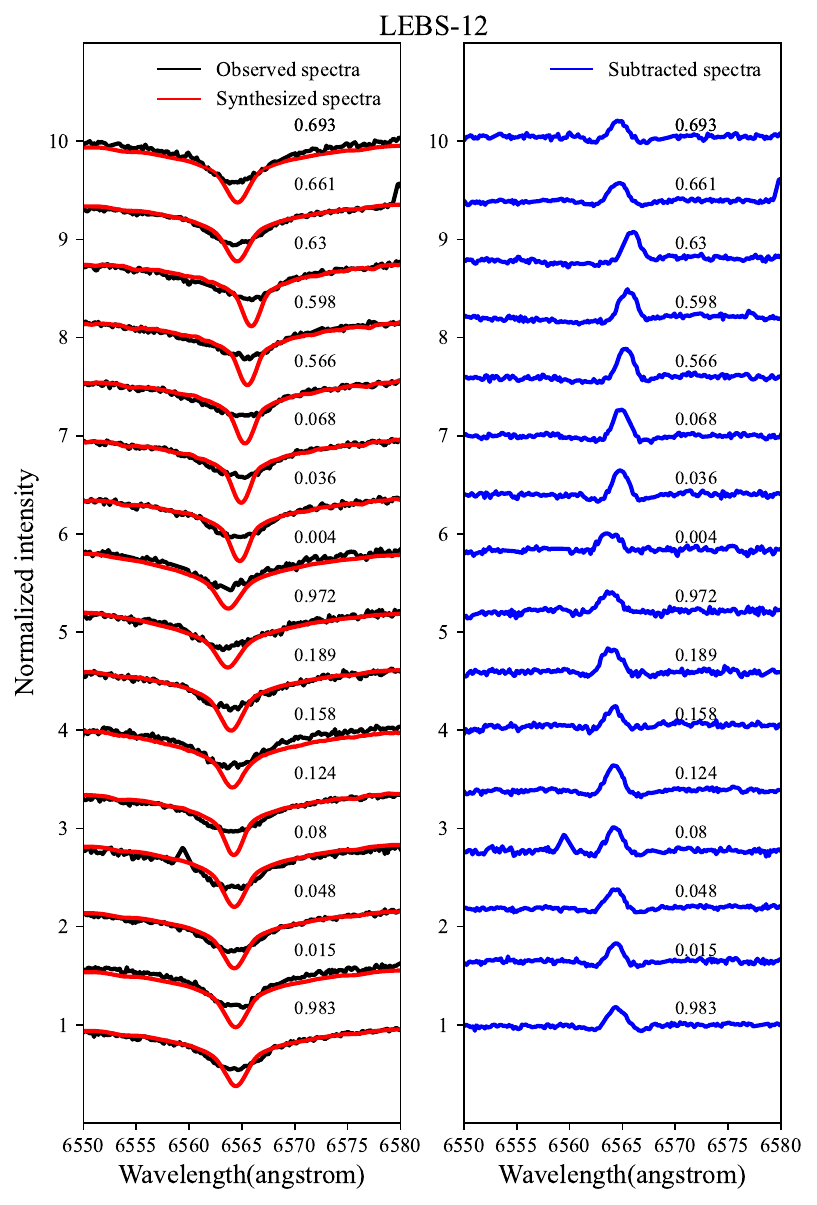}
		\vspace{0.4cm}  
		\caption{Observed, synthetic, and subtracted spectra from LAMOST medium-resolution for 13 EBs. The complete figure in the online version of this article.}
		\label{Fig 4}
	\end{figure*}

\begin{figure*}[htp]
		\centering
		\includegraphics[width=6.5cm]{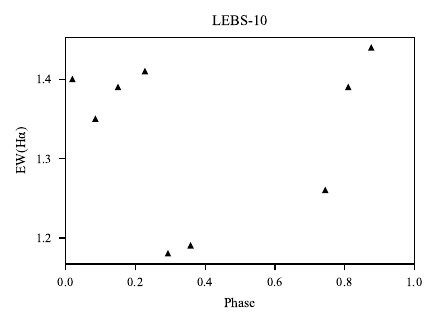}
		\hspace{1cm} 
		\includegraphics[width=6.5cm]{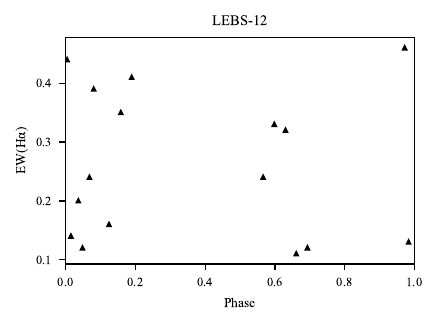}
		
		\caption{Distribution of equivalent widths (EW) of H$\alpha$ emission lines versus phases for 13 EBs. The complete figure in the online version of this article.}
		\label{Fig 5}
	\end{figure*}

	The observed, synthetic, and subtracted spectra are shown in Figure \ref{Fig 4}. The values of EW are shown in Table 2. The distribution of the EW versus phases are shown in Figure \ref{Fig 5}.

	\begin{table*}[htp!] 
		\huge  
		\begin{center} 
			\caption{The fitting parameters based on the O-C analysis and the material transfer parameters.} 
			\label{Tab 5}
			\setlength{\tabcolsep}{0.7cm} 
			\resizebox{\textwidth}{7cm}{ 
				\renewcommand\arraystretch{0.6} 
				\begin{tabular}{ccccccc}  
					
					\hline
					\multicolumn{1}{l}{Target} & $P_{C}$ (days)      & $T_{C}$ (days)   & $\beta$ (d/yr)    & dM$_{1}$/dt (M$_{\odot}$/yr) & $\tau$ (yr)    & $\tau_{th}$ (yr)   \\ 
					\hline    
					LEBS-01    & 0.6244520(±3.00$\times10^{-8}$)   & 2459825.47905(±0.00016) & —                              & —                      & —                    & —                    \\ 
					LEBS-03    & 2.3950521(±5.60$\times10^{-7}$)   & 2459825.66295(±0.00062) & —                              & —                      & —                    & —                    \\ 
					LEBS-04    & 0.4191665(±5.61$\times10^{-8}$)   & 2459825.70845(±0.00011) & -1.26(±0.00)$\times10^{-9}$  & -2.35$\times10^{-7}$ & 4.22$\times10^{6}$ & 3.56$\times10^{6}$ \\ 
					LEBS-05    & 0.6076468(±3.04$\times10^{-7}$)   & 2459825.41955(±0.00028) & 9.06(±0.60)$\times10^{-10}$  & 7.42$\times10^{-8}$  & —                    & —                    \\ 
					LEBS-06    & 2.3843446(±7.65$\times10^{-7}$)   & 2459825.37521(±0.00091) & —                              & —                      & —                    & —                    \\ 
					LEBS-07    & 0.6009719(±1.71$\times10^{-7}$)   & 2459825.45499(±0.00016) & 5.54(±0.33)$\times10^{-10}$  & 1.50$\times10^{-7}$  & —                    & —                    \\ 
					LEBS-08    & 0.3721811(±6.97$\times10^{-8}$)   & 2459825.50811(±0.00012) & 2.50(±0.08)$\times10^{-10}$  & 1.35$\times10^{-7}$  & —                    & —                    \\ 
					LEBS-09    & 2.2746678(±2.27$\times10^{-6}$)   & 2459827.54936(±0.00107) & -3.60(±1.38)$\times10^{-9}$  & -3.65$\times10^{-8}$ & 9.03$\times10^{7}$ & 5.07$\times10^{6}$ \\ 
					LEBS-10    & 0.2470799(±1.00$\times10^{-8}$)   & 2459825.34987(±0.00011) & —                              & —                      & —                    & —                    \\ 
					LEBS-11    & 0.4339908(±1.93$\times10^{-7}$)   & 2459825.66162(±0.00021) & -7.30(±0.28)$\times10^{-10}$ & -2.03$\times10^{-9}$ & 3.69$\times10^{6}$ & 1.19$\times10^{6}$ \\ 
					LEBS-12    & 0.5143119(±3.00$\times10^{-8}$)   & 2459825.48370(±0.00009) & —                              & —                      & —                    & —                    \\ 
					LEBS-16    & 0.3182739(±3.20$\times10^{-7}$)   & 2459464.26661(±0.00108) & -3.78(±3.24)$\times10^{-11}$ & -8.63$\times10^{-7}$ & 1.16$\times10^{6}$ & 4.04$\times10^{6}$ \\ 
					LEBS-20    & 0.3457552(±9.23$\times10^{-8}$)   & 2459825.47218(±0.00016) & -2.62(±0.10)$\times10^{-10}$ & -5.53$\times10^{-7}$ & 1.79$\times10^{6}$ & 1.16$\times10^{7}$ \\ 
					LEBS-22    & 0.4003590(±2.63$\times10^{-7}$)   & 2459825.48997(±0.00038) & -5.04(±0.34)$\times10^{-10}$ & -1.27$\times10^{-7}$ & 1.13$\times10^{7}$ & 3.00$\times10^{7}$ \\ 
					LEBS-24    & 0.3970362(±4.53$\times10^{-6}$)   & 2459825.66896(±0.02797) & -1.55(±0.13)$\times10^{-10}$ & -8.50$\times10^{-8}$ & 1.18$\times10^{7}$ & 3.00$\times10^{7}$ \\ 
					LEBS-25    & 0.4368378(±1.72$\times10^{-7}$)   & 2459825.76771(±0.00016) & 3.18(±0.26)$\times10^{-10}$  & 9.69$\times10^{-8}$  & —                    & —                    \\ 
					LEBS-26(P) & 21.0717501(±1.09$\times10^{-6}$) & 2459832.63337(±0.00125) & —                              & —                      & —                    & —                    \\ 
					LEBS-26(S) & 21.0717467(±1.19$\times10^{-6}$) & 2459838.35176(±0.00083) & —                              & —                      & —                    & —                    \\ 
					LEBS-28    & 0.4167704(±2.74$\times10^{-6}$)   & 2459825.63610(±0.00358) & -6.62(±3.74)$\times10^{-10}$ & -1.93$\times10^{-7}$ & 7.14$\times10^{6}$ & 5.01$\times10^{7}$ \\ 
					LEBS-29    & 5.1008655(±7.34$\times10^{-6}$)   & 2459827.46437(±0.00561) & —                              & —                      & —                    & —                    \\ 
					LEBS-33    & 0.5325326(±7.00$\times10^{-8}$)   & 2459825.76391(±0.00012) & —                              & —                      & —                    & —                    \\ 
					LEBS-34    & 0.3933864(±1.58$\times10^{-7}$)   & 2459825.37641(±0.00016) & -3.72(±0.21)$\times10^{-10}$ & -6.19$\times10^{-8}$ & 1.96$\times10^{7}$ & 1.45$\times10^{7}$ \\ 
					LEBS-39    & 0.3906593(±2.25$\times10^{-7}$)   & 2459826.39745(±0.00021) & 5.96(±0.30)$\times10^{-10}$  & 3.34$\times10^{-7}$  & —                    & —                    \\ 
					LEBS-42    & 0.2669553(±4
			\end{tabular}}
			\begin{tablenotes}    
				\footnotesize           
				\item[1] "$P_{c}$ (days)" and "$T_{c}$ (days)" represent the corrected value of the initial epoch and the orbital period, respectively. "$\beta$ (d/yr)" represents the long-term changing rate of the orbital period. "dM$_{1}$/dt (M$_{\odot}$/yr)" represents the rate of material transfer.  "$\tau$ (yr)" and "$\tau_{th}$ (yr)" represent the material transfer time-scale and thermal time-scale, respectively. "A", "$P_{mod}$", and "$\varphi$" represent the quasi-sinusoidal change's amplitude, period, and initial phase.         
			\end{tablenotes}       
		\end{center}
	\end{table*}
	
	\section{Discusstions and Conclusions}
	We detected 53 variable stars by analyzing photometric data observed by the 10 cm telescope for 13 days and by the $\emph{TESS}$ sky survey. 48 are EBs, and 2 are newly discovered. Multi-band light curves and radial velocity curves were used to determine the orbital parameters and absolute parameters of 11 targets. For 35 targets without radial velocity curves and 2 targets with unavailable radial velocity curves, we used the q-search method or the temperature ratio method to determine the initial mass ratios and obtained their orbital parameters using the W-D Code. Based on the orbital parameters of 48 EBs, 15 EBs belonged to detached systems, 1 to semi-detached systems, and 32 to contact systems LEBS-06 is a detached system with spots on both components and exhibits double-peaked H$\alpha$ emission lines. In binary systems, double-peaked H$\alpha$ emission lines are typically the result of either an accretion disk or chromospheric activity. If the double-peaked H$\alpha$ emission lines are caused by the accretion disk, they would generally display a broader velocity range and a more symmetrical profile \citep{a127}, with the position of the emission lines remaining constant across all observed phases \citep{a121}. However, the double-peaked H$\alpha$ emission lines of LEBS-06 only appear around phases 0.25 and 0.75, with an asymmetric profile, and their position is variable. Therefore, we believe that the double-peaked H$\alpha$ emission lines of LEBS-06 are due to the intense chromospheric activity of the two components. LEBS-21, LEBS-22, LEBS-28, LEBS-34, LEBS-41, and LEBS-48 are totally eclipsing contact systems, indicating their orbital parameters determined only by photometric data are reliable \citep{a31,a70}. A combined analysis of photometric and spectroscopic observations by \cite{a120} indicated that LEBS-04 was an overcontact system, with a mass ratio of 0.243(±0.008), a contact degree of 32.1\%, and a orbital inclination of 62.214(±0.368)$^\circ$. Our study shows that its mass ratio is 0.214(±0.002), its orbital inclination is 61.645(±0.088)$^\circ$, and its contact degree is 41.0(±4.4)\%. \cite{a78} calculated the orbit parameters of LEBS-20 (q=1.256(±0.004), i=80.37(±0.5)$^\circ$, f=2.6\%). Our work obtained the similar results (q=1.474(±0.012), i=80.463(±0.073)$^\circ$, f=2.7(±3.1)\%). Moreover, the light curves in our work showed a weak O’Connell effect, so we added a spot on its primary star. Given the impact of different precision in observational data and the impact of the starspots, our research results are basically consistent with those determined by previous studies. LEBS-31 had a W UMa-type light curve, but the W-D result suggested that it was a detached system. A high fill-out factor  (98.0, 98.5\%) for the primary and secondary components indicated that it may evolve into a contact system in the near future. LEBS-18 had a similar fill-out factor (77.3, 99.3\%) of their primary and secondary components. LEBS-40 showed a $\beta$ Lyrae light curve, but the W-D result indicated that it is a contact binary with a contact degree of 7.2\%, implying that it is a newly formed contact system.
	
	\subsection{Absolute physical parameters}
	
	Absolute physical parameters help to study the formation, structure, evolution, and ultimate fate of EBs \citep{a114,a70}. The absolute physical parameters of the 11 targets with radial velocity curves can be obtained directly by W-D results. For 35 targets without radial velocity curves and 2 targets with unavailable radial velocity curves, assuming that the more massive components of each target are main-sequence stars and primary components, based on the temperature-to-mass relationship \citep{a100} and the mass ratios, the masses of the two components were calculated. According to Kepler's third law $(M_{1}+M_{2}=\frac{0.0134a^{3}}{p^{2}})$, we obtained the semi-major axis $\emph{a}$. The radius $R_{1,2}$ was given by $R_{1,2}=a\times r_{1,2}$. Then, we used $L_{1,2} = (R_{1,2}/R_{\odot})^{2} (T_{1,2}/T_{\odot})^{4}$ to obtain the luminosity $L_{1,2}$. The errors of the above parameters were calculated with the error transfer formula. The absolute parameters of all targets are shown in Table 6.
	
	\subsection{Orbital period variations}
	\subsubsection{Long-term increase or decrease}
	
	There are many reasons for orbital period variations. Typically, the long-term increase in the period can be explained by material transfer from the less massive component to the more massive one. At the same time, the long-term decrease in orbital period can be explained by material transfer from the more massive component to the less one, angular momentum loss, or both. If the period variation is caused by material transfer, assuming that the angular momentum and total mass are constant, the rate of material transfer can be calculated by the equation \citep{a142}
	\begin{equation}\label{eq6}
		\frac{\dot{P} }{P} = -3\dot{M_{1}} (\frac{1}{M_{1}} -\frac{1}{M_{2}}) .
	\end{equation}
	To further determine the reason for the long-term decrease in orbital period, the material transfer time-scale $\tau$ and the thermal time-scale $\tau_{th}$ were calculated by the equation
	\begin{equation}\label{eq7}
		\tau\sim \frac{M_{1}}{\dot{M_{1}} }  ,  \tau_{th}=\frac{GM_{1}^{2}}{R_{1}L_{1}} .
	\end{equation}
	The rate of material transfer $\dot{M_{1}}$, the mass transfer time-scale $\tau$, and the thermal time-scale $\tau_{th}$ are shown in Table 5. As can be seen, for LEBS-04, LEBS-11, LEBS-16, LEBS-22, LEBS-24, LEBS-34, and LEBS-42, the time-scale $\tau$ and the thermal time-scale $\tau_{th}$ are within the same order of magnitude, implying that the long-term decrease in orbital period is caused by material transfer. For LEBS-09, LEBS-20, LEBS-28, and LEBS-47, the time-scale $\tau$ and the thermal time-scale $\tau_{th}$ are not within the same order of magnitude, implying that the long-term decrease in orbital period is attributed to the loss of orbital angular momentum or a combination of both.
	
	\subsubsection{Possible mechanisms of the cyclic oscillation}
	
	The O - C curves of LEBS-14 and LEBS-41 showed a quasi-periodic variation. For LEBS-14, $P_{mod}$ = 20.72(±1.12) yr and an amplitude of A = 0.07(±0.01) days. For LEBS-41, $P_{mod}$ = 32.52(±21.68) yr and an amplitude of A = 0.02(±0.03) days. Because the amplitude of the periodic oscillation for LEBS-41 is consistent with zero, the reason for its periodic oscillation is not discussed. Currently, cyclic oscillation can be caused by the magnetic activity of one or both components \citep{a106} or the light travel time effect (LTTE) via a third body \citep{a113, a75}.

	\begin{table*}[htp!]
		\large   
		\label{Tab 6}
		\caption{Absolute parameters of 48 EBs. } 
		\setlength{\tabcolsep}{3mm} 
		\resizebox{\textwidth}{11cm}{   
			\renewcommand\arraystretch{1.06}  
			\centering
			\begin{tabular}{cccccccc}  
				\hline
				\multicolumn{1}{l}{Target} & $M_{1}(M_{\odot})$ & $M_{2}(M_{\odot})$  & $a(R_{\odot})$   & $R_{1}(R_{\odot})$ & $R_{2}(R_{\odot})$ & $L_{1}(L_{\odot})$ & $L_{2}(L_{\odot})$ \\ \hline          
				LEBS-01 & 1.303(±0.108) & 0.267(±0.025) & 3.524(±0.103)  & 1.931(±0.058) & 0.980(±0.039) & 7.910(±0.611)   & 1.259(±0.146)  \\
				LEBS-02 & 2.600(±0.131) & 1.658(±0.108) & 7.407(±0.139)  & 3.176(±0.082) & 2.593(±0.104) & 20.080(±1.547)  & 11.483(±1.773) \\
				LEBS-03 & 1.298(±0.189) & 0.587(±0.080) & 9.306(±0.444)  & 1.604(±0.087) & 1.408(±0.070) & 2.418(±0.360)   & 2.033(±0.247)  \\
				LEBS-04 & 0.993(±0.089) & 0.212(±0.021) & 2.509(±0.077)  & 1.346(±0.044) & 0.695(±0.038) & 2.773(±0.325)   & 0.680(±0.144)  \\
				LEBS-05 & 1.062(±0.040) & 0.201(±0.008) & 3.264(±0.042)  & 1.753(±0.024) & 0.841(±0.019) & 6.198(±0.226)   & 0.935(±0.062)  \\
				LEBS-06 & 0.710(±0.011) & 0.653(±0.015) & 8.329(±0.053)  & 0.838(±0.007) & 0.745(±0.007) & 0.424(±0.051)   & 0.277(±0.056)  \\
				LEBS-07 & 1.353(±0.064) & 0.505(±0.028) & 3.685(±0.061)  & 1.854(±0.035) & 1.235(±0.042) & 6.179(±0.335)   & 2.850(±0.300)  \\
				LEBS-08 & 0.932(±0.010) & 0.370(±0.001) & 2.377(±0.007)  & 1.114(±0.011) & 0.733(±0.004) & 1.038(±0.152)   & 0.575(±0.049)  \\
				LEBS-09 & 3.295(±0.125) & 0.381(±0.018) & 11.234(±0.146) & 4.139(±0.065) & 2.403(±0.153) & 70.647(±5.172)  & 4.931(±1.008)  \\
				LEBS-10 & 0.780(±0.082) & 0.350(±0.036) & 1.726(±0.060)  & 0.793(±0.030) & 0.552(±0.020) & 0.217(±0.039)   & 0.128(±0.017)  \\
				LEBS-11 & 1.330(±0.019) & 0.848(±0.085) & 5.927(±0.002)  & 2.582(±0.005) & 1.882(±0.003) & 13.54(±0.250)   & 2.031(±0.075)  \\
				LEBS-12 & 1.610(±0.110) & 0.110(±0.001) & 3.238(±0.001)  & 2.040(±0.003) & 0.664(±0.048) & 9.831(±0.327)   & 0.988(±0.235)  \\
				LEBS-13 & 0.940(±0.040) & 0.240(±0.004) & 2.159(±0.002)  & 1.127(±0.018) & 0.637(±0.004) & 1.014(±0.078)   & 0.401(±0.013)  \\
				LEBS-14 & 1.030(±0.030) & 0.200(±0.001) & 2.632(±0.001)  & 1.395(±0.007) & 0.655(±0.002) & 2.059(±0.386)   & 0.973(±0.099)  \\
				LEBS-15 & 1.380(±0.050) & 0.760(±0.012) & 2.854(±0.005)  & 1.262(±0.007) & 0.953(±0.024) & 2.817(±0.164)   & 1.418(±0.218)  \\
				LEBS-16 & 1.000(±0.010) & 0.950(±0.106) & 2.452(±0.044)  & 1.018(±0.152) & 0.954(±0.042) & 1.019(±0.404)   & 1.065(±0.137)  \\
				LEBS-17 & 1.080(±0.020) & 1.060(±0.004) & 35.800(±0.024) & 1.038(±0.360) & 0.752(±0.360) & 1.221(±0.886)   & 0.399(±0.490)  \\
				LEBS-18 & 1.750(±0.140) & 0.260(±0.204) & 5.614(±0.190)  & 2.234(±0.108) & 1.325(±0.056) & 12.951(±1.637)  & 2.148(±0.465)  \\
				LEBS-19 & 1.330(±0.080) & 1.180(±0.047) & 14.700(±0.092) & 1.705(±0.018) & 0.941(±0.013) & 4.637(±0.313)   & 1.169(±0.146)  \\
				LEBS-20 & 0.990(±0.010) & 0.670(±0.005) & 2.456(±0.003)  & 1.017(±0.010) & 0.850(±0.003) & 1.007(±0.081)   & 0.728(±0.030)  \\
				LEBS-21 & 0.880(±0.020) & 0.250(±0.005) & 2.195(±0.003)  & 1.089(±0.007) & 0.613(±0.023) & 0.843(±0.169)   & 0.332(±0.052)  \\
				LEBS-22 & 1.440(±0.020) & 0.270(±0.011) & 2.735(±0.006)  & 1.499(±0.010) & 0.722(±0.071) & 4.226(±0.526)   & 0.993(±0.246)  \\
				LEBS-23 & 0.700(±0.010) & 0.360(±0.003) & 1.782(±0.002)  & 0.784(±0.006) & 0.577(±0.003) & 0.202(±0.043)   & 0.149(±0.016)  \\
				LEBS-24 & 1.000(±0.030) & 0.330(±0.008) & 2.216(±0.005)  & 1.079(±0.004) & 0.663(±0.004) & 1.195(±0.058)   & 0.575(±0.015)  \\
				LEBS-25 & 1.460(±0.040) & 0.360(±0.008) & 2.959(±0.005)  & 1.489(±0.008) & 0.784(±0.035) & 4.401(±0.261)   & 1.163(±0.230)  \\
				LEBS-26 & 0.980(±0.005) & 0.850(±0.001) & 39.287(±0.007) & 0.864(±0.045) & 0.825(±0.045) & 0.693(±0.108)   & 0.595(±0.163)  \\
				LEBS-27 & 1.060(±0.020) & 0.190(±0.024) & 2.174(±0.014)  & 1.157(±0.031) & 0.587(±0.179) & 1.486(±0.092)   & 0.382(±0.278)  \\
				LEBS-28 & 1.380(±0.060) & 0.320(±0.002) & 2.804(±0.001)  & 1.477(±0.003) & 0.788(±0.011) & 3.921(±0.147)   & 1.021(±0.098)  \\
				LEBS-29 & 1.440(±0.020) & 1.400(±0.002) & 17.668(±0.004) & 1.343(±0.007) & 1.908(±0.006) & 3.370(±0.239)   & 6.521(±0.739)  \\
				LEBS-30 & 1.880(±0.050) & 1.450(±0.002) & 8.696(±0.002)  & 2.113(±0.032) & 1.800(±0.027) & 17.8127(±1.946) & 2.309(±0.524)  \\
				LEBS-31 & 0.970(±0.020) & 0.180(±0.001) & 1.777(±0.001)  & 0.945(±0.002) & 0.434(±0.002) & 0.785(±0.014)   & 0.145(±0.014)  \\
				LEBS-32 & 0.820(±0.040) & 0.710(±0.042) & 4.871(±0.045)  & 0.779(±0.021) & 0.794(±0.039) & 0.369(±0.087)   & 0.339(±0.144)  \\
				LEBS-33 & 1.500(±0.110) & 0.290(±0.062) & 3.359(±0.039)  & 1.565(±0.028) & 0.893(±0.018) & 5.368(±0.379)   & 0.230(±0.034)  \\
				LEBS-34 & 1.210(±0.040) & 0.180(±0.001) & 2.523(±0.001)  & 1.458(±0.009) & 0.691(±0.027) & 2.995(±0.111)   & 0.695(±0.108)  \\
				LEBS-35 & 0.620(±0.030) & 0.610(±0.001) & 2.541(±0.000)  & 0.363(±0.001) & 0.231(±0.002) & 0.029(±0.003)   & 0.015(±0.001)  \\
				LEBS-36 & 1.980(±0.070) & 0.660(±0.090) & 8.819(±0.100)  & 1.676(±0.085) & 2.372(±0.587) & 16.219(±2.441)  & 4.788(±3.083)  \\
				LEBS-37 & 1.460(±0.020) & 0.230(±0.015) & 11.614(±0.033) & 1.521(±0.016) & 0.743(±0.027) & 4.449(±0.381)   & 0.804(±0.176)  \\
				LEBS-38 & 0.880(±0.020) & 0.440(±0.038) & 2.024(±0.019)  & 0.921(±0.091) & 0.721(±0.028) & 0.579(±0.185)   & 0.370(±0.039)  \\
				LEBS-39 & 1.180(±0.050) & 0.440(±0.003) & 2.642(±0.002)  & 1.245(±0.009) & 0.790(±0.003) & 1.965(±0.408)   & 0.894(±0.091)  \\
				LEBS-40 & 0.900(±0.020) & 0.270(±0.002) & 2.314(±0.001)  & 1.136(±0.007) & 0.653(±0.002) & 0.958(±0.131)   & 1.012(±0.066)  \\
				LEBS-41 & 1.030(±0.030) & 0.250(±0.003) & 2.469(±0.002)  & 1.289(±0.006) & 0.701(±0.023) & 1.775(±0.105)   & 0.477(±0.081)  \\
				LEBS-42 & 0.780(±0.040) & 0.340(±0.003) & 1.813(±0.001)  & 0.834(±0.007) & 0.566(±0.001) & 0.349(±0.085)   & 0.238(±0.033)  \\
				LEBS-43 & 0.780(±0.040) & 0.500(±0.011) & 1.812(±0.005)  & 0.790(±0.021) & 0.641(±0.006) & 0.320(±0.112)   & 0.231(±0.041)  \\
				LEBS-44 & 0.700(±0.030) & 0.590(±0.044) & 4.190(±0.047)  & 0.629(±0.037) & 0.884(±0.022) & 0.147(±0.023)   & 0.428(±0.024)  \\
				LEBS-45 & 0.780(±0.040) & 0.280(±0.016) & 1.922(±0.010)  & 0.924(±0.057) & 0.603(±0.017) & 0.456(±0.070)   & 0.273(±0.016)  \\
				LEBS-46 & 1.250(±0.080) & 0.580(±0.024) & 2.747(±0.012)  & 1.327(±0.017) & 0.970(±0.066) & 2.660(±0.299)   & 1.410(±0.439)  \\
				LEBS-47 & 1.440(±0.020) & 0.700(±0.011) & 4.997(±0.009)  & 2.234(±0.011) & 1.599(±0.040) & 9.472(±0.320)   & 3.395(±0.368)  \\
				LEBS-48 & 0.880(±0.020) & 0.210(±0.003) & 2.004(±0.002)  & 1.026(±0.014) & 0.531(±0.003) & 0.752(±0.100)   & 0.246(±0.017)             \\ \hline
		\end{tabular}}
		\begin{tablenotes}    
			\footnotesize           
			\item[1] Assuming that the more massive components are the primary components. "$M_{1}$$(M_{\odot})$", "$R_{1}$$(R_{\odot})$", and "$L_{1}$$(L_{\odot})$" represent the mass, radius, and luminosity of the primary component. "$M_{2}$$(M_{\odot})$", "$R_{2}$$(R_{\odot})$", and "$L_{2}$$(L_{\odot})$" represent the mass, radius, and luminosity of the secondary component. "a" represents the semi-major axis.
		\end{tablenotes}       
	\end{table*}

	If the cyclic modulation is caused by magnetic activity, the variation in the magnetic quadruple moment  ($\Delta Q$) can be calculated according to the following equation by Applegate (\citeyear{a106})
	\begin{equation}\label{eq8}
		\frac{\Delta P}{P} = \frac{2\pi \times A}{P_{mod} } = -9  (\frac{R}{a} )^{2}  \frac{\Delta Q}{M R^{2} } ,
	\end{equation}
	where $\emph{M}$, $\emph{R}$, and $\emph{a}$ are mass, radii of the active component, and the semi-major axis of the targets. For LEBS-14,  the values of $\Delta Q$ were calculated as, $\Delta Q_{1}=3.350(\pm0.126) \times 10^{50}\, g \,cm^{2}$ and $\Delta Q_{2}=6.472(\pm0.244) \times 10^{49}\, g \,cm^{2}$. These results are clearly smaller than the typical values of  $10^{51} \sim  10^{52}\, g \,cm^{2}$ for close binaries \citep{a107}. Therefore, the Applegate mechanism is not reasonable to be used to explain the periodic variations.
	
	Another possible mechanism is the LTTE. The distance of the binary system to the barycenter of the triple system was calculated with the equation
	\begin{equation}\label{eq9}
		a_{12}\sin i_{3}=A\times c,
	\end{equation}
	where $\emph{A}$ is the amplitude of the O - C oscillation and c is the speed of the light. We derived $a_{12}\sin i_{3}$ = 11.919(±1.093) AU. The mass function of the third body $f(M_{3})$ was calculated by the follow equation
	
	\begin{equation}\label{eq10}
		f(M_{3} ) = \frac{(M_{3}\sin i_{3})^{3} }{(M_{1}+M_{2}+M_{3})^{2} } = \frac{4\pi^{2}}{GP_{3}^{2}}\times (a_{12}\sin i_{3})^3 ,
	\end{equation}
	the mass function of the third body was determined as $f(M_{3})$ = 3.943(±1.085) $M_{\odot}$. When the orbital inclination of the third body ($i_{3}$) was 90$^\circ$, the minimum mass of the third body was determined to be $M_{3min}$ = 5.570(±1.953) $M_{\odot}$, which suggests that it is likely a black hole. This makes LEBS-14 a valuable subject for study, necessitating long-term observational follow-up to confirm the current conclusions of the O-C analysis.

	\subsection{Evolutionary state}
	
\begin{figure*}[ht!]
		\centering
		\includegraphics[width=8.5cm]{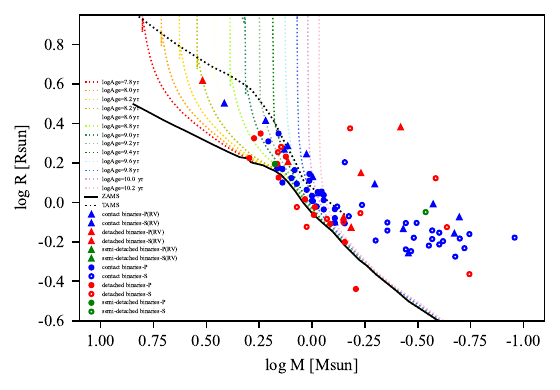}
		\vspace{0.4cm}
		\includegraphics[width=8.5cm]{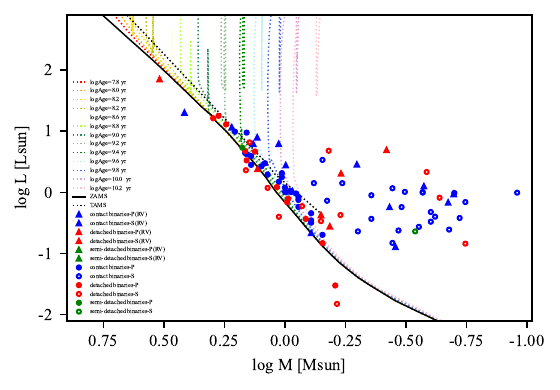}
		\caption{The M-R and M-L distributions for EBs. The colored dotted lines indicate the evolutionary tracks for solar chemical compositions. The solid and dashed lines indicate the zero-age main sequence (ZAMS) and the terminal-age main sequence (TAMS), respectively. The triangular and circular symbols denote targets with and without radial velocity curves, respectively. The solid and hollow symbols denote the more and less massive components, respectively. The red, blue, and green symbols denote detached, contact, and semi-detached systems, respectively.}
		\label{Fig 6}
	\end{figure*}

	To study the evolutionary state of the EBs, we plotted their mass-luminosities (M-L) and mass-radius (M-R) distributions as in Figure \ref{Fig 6}, in which evolutionary tracks for solar chemical compositions (as colored dotted lines) are taken from \cite{a145}, the zero-age main sequence (ZAMS) (as solid lines) and the terminal-age main sequence (TAMS) (as dashed lines) are plotted. The triangular and circular symbols denote targets with and without radial velocity curves, respectively. The solid and hollow symbols denote the more and less massive components, respectively. The red, blue, and green symbols denote detached, contact, and semi-detached systems, respectively. As shown in the figure, the more massive primary stars are closer to ZAMS than the less massive secondary stars, and the less massive secondary stars are closer to the TAMS, which means that the more massive primary stars are less evolved than secondaries. In binary systems, the initially more massive component evolves more rapidly, expanding to fill its Roche Lobe and transferring material to the less massive component. This process results in the less massive component gaining mass, while the more massive component concurrently loses mass, leading to a reversal in their mass ratio \citep{a128}. Therefore, in semi-detached and contact systems, due to material transfer and the reversal of the mass ratio, the initial less massive component becomes the more massive one and is now at the main sequence stage. Conversely, the initial more massive component has essentially evolved off the main sequence stage. For detached systems with rapidly evolving secondary stars, we think it may be because they are currently in the stage of contact broken stage during Thermal Relaxation Oscillations (TRO) \citep{a136,a137,a129}, which is also caused by a previous reversal in the mass ratio. As shown in Figure 6, the less massive components are located outside the evolutionary models, not because they have evolved to this position, but because the material transfer between the two components, which results in the less massive components exhibiting over-luminosity and over-size relative to main sequence stars of the same mass.
	
\begin{figure*}[ht!]
		\centering
		\includegraphics[width=9cm]{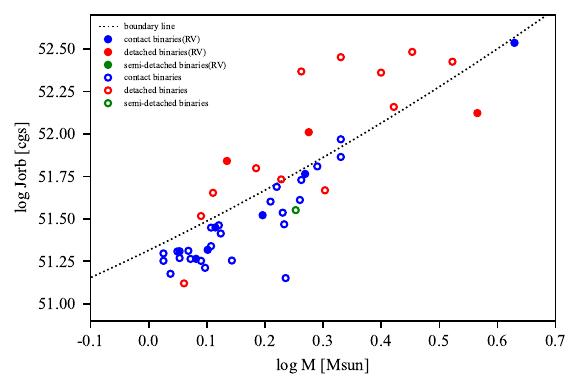}
		\vspace{0.4cm}
		
		\caption{The relation between orbital angular momentum and total mass for EBs. The dashed line indicates the boundary line between the detached and contact binaries. The solid and hollow symbols indicate targets with and without radial velocity curves, respectively. The red, blue, and green symbols denote detached, contact, and semi-detached systems, respectively.}
		\label{Fig 7}
	\end{figure*}

	Then, we calculated the orbital angular momentum $J_{orb}$  using the following equation \citep{a72}:
	\begin{equation}\label{eq8}
		J_{orb} = \frac{q}{(1+q)^{2}} \sqrt[3]{\frac{G^{2}}{2\pi}M_{T}^{5}P} 
	\end{equation}
	where $M_{T}$ is the total mass of the binary, $\emph{P}$ is the orbital period, and $\emph{q}$ is the mass ratio. The total mass versus orbital angular momentum distribution is plotted in Figure \ref{Fig 7}. The dashed line indicates the boundary line between the detached and contact binaries \citep{a72}. The solid and hollow symbols indicate targets with and without radial velocity curves, respectively. The red, blue, and green symbols denote detached, contact, and semi-detached systems, respectively. Almost all contact binaries are below the boundary line, representing a smaller orbital angular momentum for the contact binary with the same total mass. The reason for this phenomenon is the angular momentum loss due to mass loss or magnetic stellar wind during the formation of contact binaries. This conclusion is consistent with previous studies (e.g., \citeauthor{a70} \citeyear{a70}).
	
	In conclusion, 48 EBs were detected around the field, RA: $23^h$$01^m$$51.00^s$, Dec: +34$^\circ$36$^\prime$45$^{\prime \prime}$. The orbital parameters were determined, with 15 EBs belonging to detached systems, 1 to semi-detached systems, and 32 to contact systems. The orbital period variation was analyzed using the O-C method. Additionally, chromospheric activity in the LAMOST medium-resolution spectra was analyzed. Finally, the evolutionary states were discussed.

	\section{Data Availability}
	The 10 cm telescope data and \emph{TESS} data used in this research can be downloaded through China VO (https://nadc.china-vo.org/res/r101438/).

	\begin{acknowledgements}
		
		We are grateful to the reviewer for the valuable suggestions and comments, which has significantly contributed to the enhancement of this manuscript. This work was supported by National Natural Science Foundation of China (NSFC) (No. 12273018), and the Joint Research Fund in Astronomy (No.U1931103) under cooperative agreement between NSFC and Chinese Academy of Sciences (CAS), and by the Qilu Young Researcher Project of Shandong University, and by Young Data Scientist Project of the National Astronomical Data Center and by the Cultiation Project for LAMOST Scientific Payoff and Research Achievement of CAMSCAS, and by the Chinese Academy of Science Interdisciplinary Innovation Team. The calculations in this work were carried out at Supercomputing Center of Shandong University, Weihai.

		The spectral data were provided by Guoshoujing Telescope (the Large Sky Area Multi-Object Fiber Spectroscopic Telescope LAMOST), which is a national major scientific project built by the Chinese Academy of Sciences. Funding for the project has been provided by the National Development and Reform Commission. LAMOST is operated and managed by the National Astronomical Observatories, Chinese Academy of Sciences.

		This work includes data collected by the TESS mission. Funding for the TESS mission is provided by NASA Science Mission Directorate. We acknowledge the TESS team for its support of this work.  We thank Las Cumbres Observatory and its staff for their continued support of ASAS-SN. ASAS-SN is funded in part by the Gordon and Betty Moore Foundation through grant nos. GBMF5490 and GBMF10501 to the Ohio State University, and also funded in part by the Alfred P. Sloan Foundation grant no.G-2021-14192. This paper makes use of data from ZTF. ZTF is supported by the National Science Foundation under grant no.AST-1440341 and a collaboration including Caltech, IPAC, the Weizmann Institute for Science, the Oskar Klein Center at Stockholm University, the University of Maryland, the University of Washington, Deutsches Elektronen-Synchrotron and Humboldt University, Los Alamos National Laboratories, the TANGO Consortium of Taiwan, the University of Wisconsin at Milwaukee, and Lawrence Berkeley National Laboratories. Operations are conducted by COO, IPAC, and UW. This publication makes use of data products from the AAVSO Photometric All Sky Survey (APASS). Funded by the Robert Martin Ayers Sciences Fund and the National Science Foundation. This publication makes use of data products from the Two Micron All Sky Survey, which is a joint project of the University of Massachusetts and the Infrared Processing and Analysis Center/California Institute of Technology, funded by the National Aeronautics and Space Administration and the Science Foundation. This paper makes use of data from the DR1 of the WASP data, Butters et al. (2010) as provided by the WASP consortium, and the computing and storage facilities at the CERIT Scientific Cloud, reg. no. CZ.1.05/3.2.00/08.0144, which is operated by Masaryk University, Czech Republic. This work has made use of data from the European Space Agency (ESA) mission Gaia (https://www.cosmos.esa.int/gaia), processed by the Gaia Data Processing and Analysis Consortium (DPAC,https://www.cosmos.esa.int/web/gaia/ dpac/consortium). Funding for the DPAC has been provided by national institutions, in particular the institutions participating in the Gaia Multilateral Agreement.  
	\end{acknowledgements}

\end{document}